\documentclass[twocolumn]{aastex63}
\usepackage{hyperref}
\usepackage{csvsimple}
\usepackage{amsmath}
\usepackage{colortbl}
\shorttitle{Stellar Activity/Variability Benchmarks}
\shortauthors{Zhao et al.}

\newcommand{\project}[1]{\textsl{#1}}
\newcommand{\acronym}[1]{{\small{#1}}}
\newcommand{\vald}{\project{\acronym{VALD}}}
\newcommand{\expres}{\acronym{EXPRES}}
\newcommand{\apt}{\acronym{APT}s}
\newcommand{\ldt}{\acronym{LDT}}

\newcommand{\espresso}{\acronym{ESPRESSO}}
\newcommand{\neid}{\acronym{NEID}}
\newcommand{\hires}{\acronym{HIRES}}
\newcommand{\pfs}{\acronym{PFS}}
\newcommand{\chiron}{\acronym{CHIRON}}
\newcommand{\harps}{\acronym{HARPS}}

\newcommand{\oxgen}{OxBridGen}
\newcommand{\wobble}{\project{wobble}}
\newcommand{\excalibur}{\project{excalibur}}
\newcommand{\george}{\project{george}}
\newcommand{\horatio}{Generative RR}
\newcommand{\hamlet}{Discriminative RR}
\newcommand{\selenite}{\project{\acronym{SELENITE}}}
\newcommand{\scalpels}{\project{\acronym{SCALPELS}}}
\newcommand{\glom}{\project{\acronym{GLOM}}}
\newcommand{\matern}{Mat\'{e}rn $\frac{5}{2}$}
\newcommand{\rassine}{\project{\acronym{RASSINE}}}

\newcommand{\yarara}{\project{\acronym{YARARA}}}
\newcommand{\lbl}{\project{\acronym{LBL}}}
\newcommand{\lblspec}{\lbl+PCA\textsubscript{Spec.}}
\newcommand{\lblrvel}{\lbl+PCA\textsubscript{RV}}
\newcommand{\lblboth}{\lbl+PCA\textsubscript{Spec./RV}}
\newcommand{\zlsd}{\project{\acronym{ZLSD}}}
\newcommand{\pwgp}{\project{\acronym{PWGP}}}

\newcommand{\pyaneti}{\project{pyaneti}}
\newcommand{\ccfprime}{\project{CCF Prime}}
\newcommand{\dcpca}{\project{\acronym{DCPCA}}}
\newcommand{\fiesta}{\project{\acronym{FIESTA}}}

\newcommand{\gprn}{\project{\acronym{GPRN}}}
\newcommand{\fdpca}{\project{\acronym{FDPCA}}}
\newcommand{\hgrv}{\project{\acronym{HGRV}}}
\newcommand{\safe}{\project{\acronym{SAFE}}}
\newcommand{\vspan}{V\textsubscript{span}}
\newcommand{\mleprv}{ML\_EPRVs}

\newcommand{\name}{\expres\ Stellar-Signals Project}
\newcommand{\essp}{\project{\acronym{ESSP}}}
\newcommand{\scc}[1]{SCC\textsubscript{#1}}
\newcommand{\tcet}{\mbox{$\tau~{\rm Ceti}$}} 
\newcommand{\ha}{H$\alpha$}
\newcommand{\rhk}{$\log R'_{\rm HK}$}
\newcommand{\teff}{${\rm T_{eff}}$}
\newcommand{\cms}{\mbox{cm s$^{-1}$}}
\newcommand{\ms}{\mbox{m s$^{-1}$}}
\newcommand{\kms}{\mbox{km s$^{-1}$}}

\begin{document}
\title{The \name\ II. State of the Field in Disentangling Photospheric Velocities}

\correspondingauthor{Lily Zhao}
\email{lzhao@flatironinstitute.org}

\author[0000-0002-3852-3590]{Lily L. Zhao} 
\affiliation{Department of Astronomy, Yale University, 52 Hillhouse Ave., New Haven, CT 06511, USA} 
\affiliation{Center for Computational Astrophysics, Flatiron Institute, Simons Foundation, 162 Fifth Avenue, New York, NY 10010, USA} 
\affiliation{CCA Team}

\author[0000-0003-2221-0861]{Debra A. Fischer} 
\affiliation{Department of Astronomy, Yale University, 52 Hillhouse Ave., New Haven, CT 06511, USA} 

\author[0000-0001-6545-639X]{Eric B. Ford} 
\affiliation{Department of Astronomy \& Astrophysics, 525 Davey Laboratory, The Pennsylvania State University, University Park, PA, 16802, USA} 
\affiliation{Center for Exoplanets and Habitable Worlds, 525 Davey Laboratory, The Pennsylvania State University, University Park, PA, 16802, USA} 
\affiliation{Institute for Computational \& Data Sciences, The Pennsylvania State University, University Park, PA, 16802, USA} 
\affiliation{Institute for Advanced Sciences} 
\affiliation{PennState Team} 

\author[0000-0002-5013-5769]{Alex Wise} 
\affiliation{Department of Astronomy \& Astrophysics, 525 Davey Laboratory, The Pennsylvania State University, University Park, PA, 16802, USA} 
\affiliation{PennState Team} 

\author[0000-0002-2207-0750]{Micha\"{e}l Cretignier} 
\affiliation{Astronomy Department of the University of Geneva, 51 Chemin de Pegasi 51, 1290 Versoix, Switzerland} 
\affiliation{Geneva Team} 

\author[0000-0003-1453-0574]{Suzanne Aigrain} 
\affiliation{Sub-department of Astrophysics, Department of Physics, University of Oxford, Oxford OX1 3RH, UK} 
\affiliation{OxBridGen Team}
 
\author[0000-0003-0563-0493]{Oscar Barragan} 
\affiliation{Sub-department of Astrophysics, Department of Physics, University of Oxford, Oxford OX1 3RH, UK} 
\affiliation{OxBridGen Team}
 
\author[0000-0001-9907-7742]{Megan Bedell} 
\affiliation{Center for Computational Astrophysics, Flatiron Institute, Simons Foundation, 162 Fifth Avenue, New York, NY 10010, USA} 
\affiliation{CCA Team}
 
\author[0000-0003-1605-5666]{Lars A. Buchhave} 
\affiliation{DTU Space, National Space Institute, Technical University of Denmark, Elektrovej 328, DK-2800 Kgs. Lyngby, Denmark} 
\affiliation{OxBridGen Team}
 
\author[0000-0001-5121-5560]{Jo\~{a}o D. Camacho} 
\affiliation{Instituto de Astrof\'{i}sica e Ci\^{e}ncias do Espa\c{c}o, Universidade do Porto, CAUP, Rua das Estrelas, 4150-762, Porto, Portugal} 
\affiliation{Departamento de F\'{i}sica e Astronomia, Faculdade de Ci\^{e}ncias, Universidade do Porto, Rua do Campo Alegre, 687, 4169-007 Porto, Portugal}
\affiliation{Porto Team}
 
\author[0000-0001-8934-7315]{Heather M. Cegla} 
\affiliation{Department of Physics, University of Warwick, Gibbet Hill Road, Coventry CV4 7AL, UK}
\affiliation{Centre for Exoplanets and Habitability, University of Warwick, Coventry CV4 7AL, UK}
\affiliation{Warwick Team}
 
\author[0000-0002-9656-2272]{Jessi Cisewski-Kehe} 
\affiliation{Department of Statistics, University of Wisconsin-Madison, 1300 University Ave., Madison, WI 53706, USA} 
 
\author[0000-0002-8863-7828]{Andrew Collier Cameron} 
\affiliation{University of St Andrews, Centre for Exoplanet Science, SUPA, School of Physics \& Astronomy, North Haugh, St Andrews KY16 9SS, UK} 
\affiliation{St.\ Andrews Team}
 
\author[0000-0002-7564-6047]{Zoe L. de Beurs} 
\affiliation{Department of Earth, Atmospheric, and Planetary Sciences, Massachusetts Institute of Technology,77 Massachusetts Avenue, 54-918, Cambridge, MA 02139}
\affiliation{Department of Astronomy, University of Texas at Austin, 2515 Speedway, Austin, Texas 78712, USA} 
\affiliation{\mleprv\ Team}
 
\author[0000-0002-8796-4974]{Sally Dodson-Robinson} 
\affiliation{Department of Physics and Astronomy, University of Delaware, 217 Sharp Lab, Newark, DE 19716, USA} 
\affiliation{Bartol Research Institute, Sharp Lab, 104 The Green, Newark, DE, 19716, USA} 
\affiliation{Sidera Team}
 
\author[0000-0002-9332-2011]{Xavier Dumusque} 
\affiliation{Astronomy Department of the University of Geneva, 51 Chemin de Pegasi 51, 1290 Versoix, Switzerland} 
\affiliation{Geneva Team}
 
\author[0000-0002-6728-244X]{Jo\~{a}o P. Faria} 
\affiliation{Instituto de Astrof\'{i}sica e Ci\^{e}ncias do Espa\c{c}o, Universidade do Porto, CAUP, Rua das Estrelas, 4150-762, Porto, Portugal} 
\affiliation{Departamento de F\'{i}sica e Astronomia, Faculdade de Ci\^{e}ncias, Universidade do Porto, Rua do Campo Alegre, 687, 4169-007 Porto, Portugal}
\affiliation{Porto Team}
 
\author[0000-0002-1743-3684]{Christian Gilbertson} 
\affiliation{Department of Astronomy \& Astrophysics, 525 Davey Laboratory, The Pennsylvania State University, University Park, PA, 16802, USA} 
\affiliation{Center for Exoplanets and Habitable Worlds, 525 Davey Laboratory, The Pennsylvania State University, University Park, PA, 16802, USA} 
\affiliation{Institute for Computational \& Data Sciences, The Pennsylvania State University, University Park, PA, 16802, USA} 
\affiliation{PennState Team}
 
\author[0000-0003-3996-773X]{Charlotte Haley} 
\affiliation{Mathematics and Computer Science Division, Argonne National Laboratory, Lemont, IL} 
\affiliation{Sidera Team}
 
\author[0000-0001-6771-4583]{Justin Harrell} 
\affiliation{Department of Physics and Astronomy, University of Delaware, 217 Sharp Lab, Newark, DE 19716, USA} 
\affiliation{Sidera Team}
 
\author[0000-0003-2866-9403]{David W. Hogg} 
\affil{Center for Cosmology and Particle Physics, Department of Physics, New York University, 726 Broadway, New York, NY 10003, USA}
\affil{Center for Computational Astrophysics, Flatiron Institute, Simons Foundation, 162 Fifth Avenue, New York, NY 10010, USA}
\affil{Center for Data Science, New York University, 60 Fifth Avenue, New York, NY 10011, USA}
\affil{Max-Planck-Institut f\"{u}r Astronomie, K\"{o}nigstuhl 17, D-69117 Heidelberg, Germany}
\affiliation{CCA Team}
 
\author[0000-0001-8936-6276]{Parker Holzer} 
\affiliation{Department of Statistics and Data Science, Yale University, 24 Hillhouse Avenue, New Haven, CT 06511, USA} 
 
\author[0000-0002-1715-6939]{Ancy Anna John} 
\affiliation{University of St Andrews, Centre for Exoplanet Science, SUPA, School of Physics \& Astronomy, North Haugh, St Andrews KY16 9SS, UK} 
\affiliation{St.\ Andrews Team}
 
\author[0000-0003-0637-5236]{Baptiste Klein} 
\affiliation{Sub-department of Astrophysics, Department of Physics, University of Oxford, Oxford OX1 3RH, UK} 
\affiliation{OxBridGen Team}
 
\author[0000-0002-8815-9416]{Marina Lafarga} 
\affiliation{Department of Physics, University of Warwick, Gibbet Hill Road, Coventry CV4 7AL, UK} 
\affiliation{Warwick Team}
 
\author[0000-0003-4047-0771]{Florian Lienhard} 
\affiliation{Astrophysics Group, Cavendish Laboratory, University of Cambridge, J.J. Thomson Avenue, Cambridge CB3 0HE, UK} 
 
\author[0000-0001-7576-6703]{Vinesh Maguire-Rajpaul} 
\affiliation{Astrophysics Group, Cavendish Laboratory, University of Cambridge, J.J. Thomson Avenue, Cambridge CB3 0HE, UK} 
\affiliation{OxBridGen Team}
 
\author[0000-0001-7254-4363]{Annelies Mortier} 
\affiliation{Astrophysics Group, Cavendish Laboratory, University of Cambridge, J.J. Thomson Avenue, Cambridge CB3 0HE, UK} 
\affiliation{Kavli Institute for Cosmology, University of Cambridge, Madingley Road, Cambridge CB3 0HA, UK} 
 
\author[0000-0003-1360-4404]{Belinda Nicholson} 
\affiliation{Sub-department of Astrophysics, Department of Physics, University of Oxford, Oxford OX1 3RH, UK} 
\affiliation{OxBridGen Team}
 
\author[0000-0002-4677-8796]{Michael L. Palumbo III} 
\affiliation{Department of Astronomy \& Astrophysics, 525 Davey Laboratory, The Pennsylvania State University, University Park, PA, 16802, USA} 
\affiliation{PennState Team}
 
\author[0000-0001-8183-459X]{Victor Ramirez Delgado} 
\affiliation{Department of Physics and Astronomy, University of Delaware, 217 Sharp Lab, Newark, DE 19716, USA} 
\affiliation{Sidera Team}
 
\author{Christopher J. Shallue} 
\affiliation{Center for Astrophysics|Harvard \& Smithsonian, 60 Garden Street, Cambridge, MA 02138, USA} 
\affiliation{\mleprv\ Team}
 
\author[0000-0001-7246-5438]{Andrew Vanderburg} 
\affiliation{Department of Physics and Kavli Institute for Astrophysics and Space Research, Massachusetts Institute of Technology, 77 Massachusetts Avenue, Cambridge, MA 02139, USA} 
\affiliation{Department of Astronomy, University of Wisconsin-Madison, Madison, WI, 53706, USA} 
\affiliation{\mleprv\ Team}
 
\author[0000-0003-1572-8531]{Pedro T. P. Viana} 
\affiliation{Instituto de Astrof\'{i}sica e Ci\^{e}ncias do Espa\c{c}o, Universidade do Porto, CAUP, Rua das Estrelas, 4150-762, Porto, Portugal} 
\affiliation{Departamento de Física e Astronomia, Faculdade de Ciências, Universidade do Porto, Rua do Campo Alegre, 687, 4169-007 Porto, Portugal}
\affiliation{Porto Team}
 
\author[0000-0001-5290-2952]{Jinglin Zhao} 
\affiliation{Department of Astronomy \& Astrophysics, 525 Davey Laboratory, The Pennsylvania State University, University Park, PA, 16802, USA} 
\affiliation{PennState Team}
 
\author[0000-0001-6143-2905]{Norbert Zicher} 
\affiliation{Sub-department of Astrophysics, Department of Physics, University of Oxford, Oxford OX1 3RH, UK} 
\affiliation{OxBridGen Team}

\author[0000-0001-9749-6150]{Samuel H. C. Cabot} 
\affiliation{Department of Astronomy, Yale University, 52 Hillhouse Ave., New Haven, CT 06511, USA} 

\author[0000-0003-4155-8513]{Gregory W. Henry} 
\affiliation{Center of Excellence in Information Systems, Tennessee State University, Nashville, TN 37209, USA}

\author[0000-0002-9288-3482]{Rachael M. Roettenbacher} 
\affiliation{Yale Center for Astronomy and Astrophysics, Yale University, 46 Hillhouse Avenue, New Haven, CT 06511, USA} 
\affiliation{Department of Astronomy, Yale University, 52 Hillhouse Ave., New Haven, CT 06511, USA}  
 
\author[0000-0002-9873-1471]{John M. Brewer} 
\affiliation{San Francisco State University University, 1600 Holloway Ave., San Francisco, CA 94132, USA} 

\author[0000-0003-4450-0368]{Joe Llama} 
\affiliation{Lowell Observatory, 1400 W. Mars Hill Rd., Flagstaff, AZ 86001, USA} 

\author[0000-0003-2168-0191]{Ryan R. Petersburg} 
\affiliation{Department of Physics, Yale University, 217 Prospect St, New Haven, CT 06511, USA} 

\author[0000-0002-4974-687X]{Andrew E. Szymkowiak} 
\affiliation{Department of Astronomy, Yale University, 52 Hillhouse Ave., New Haven, CT 06511, USA}

\begin{abstract}\noindent
Measured spectral shifts due to intrinsic stellar variability (e.g., pulsations, granulation) and activity (e.g., spots, plages) are the largest source of error for extreme precision radial velocity (EPRV) exoplanet detection.  Several methods are designed to disentangle stellar signals from true center-of-mass shifts due to planets.  The \expres\ Stellar Signals Project (\essp) presents a self-consistent comparison of 22 different methods tested on the same extreme-precision spectroscopic data from \expres.  Methods derived new activity indicators, constructed models for mapping an indicator to the needed RV correction, or separated out shape- and shift-driven RV components.  Since no ground truth is known when using real data, relative method performance is assessed using the total and nightly scatter of returned RVs and agreement between the results of different methods.  Nearly all submitted methods return a lower RV RMS than classic linear decorrelation, but no method is yet consistently reducing the RV RMS to sub-meter-per-second levels.  There is a concerning lack of agreement between the RVs returned by different methods.  These results suggest that continued progress in this field necessitates increased interpretability of methods, high-cadence data to capture stellar signals at all timescales, and continued tests like the \essp\ using consistent data sets with more advanced metrics for method performance.  Future comparisons should make use of various well-characterized data sets---such as solar data or data with known injected planetary and/or stellar signals---to better understand method performance and whether planetary signals are preserved.
\end{abstract}

\keywords{Exoplanet detection methods (489), Radial velocity (1332), Planet hosting stars (1242), Stellar activity (1580), Spectrometers (1554)}

\section{Introduction}\label{sec:intro}
With the new generation of extreme-precision spectrographs, sub-meter-per-second radial velocity (RV) measurement precision has become achievable \citep{pepe2013, schwab2016, jurgenson2016, blackman2020, petersburg2020, mascareno2020, brewer2021, pepe2021}.  Photospheric velocities from stellar variability and activity features are now the dominant source of RV scatter.

A star's radial velocity is measured by modeling Doppler shifts in absorption lines of stellar spectra.  Different forms of stellar variability will change spectra such that lines will appear shifted, deeper/shallower, or asymmetric.  These line-shape changes can be mistaken for true center-of-mass shifts in the RV analysis.  In this way, stellar signals add errors to the resultant RV measurements and can even masquerade as periodic, false planet signals \citep[e.g.,][]{rajpaul2016}.

With instrumental RV precision better than one meter per second, we must contend with obscuring photospheric velocities that arise from stellar p-mode oscillations \citep{mayor2003, bouchy2005, kjeldsen2005, arentoft2008, chaplin2019}, granulation \citep{dravins1982, kjeldsen1995, lindegren2003, dumusque2011-01, meunier2015, cegla2018, lanza2019-03}, supergranulation \citep{rieutord2010, rincon2018, meunier2019}, and large-amplitude magnetic activity features such as spots, faculae, or plages \citep{saar1997, hatzes2002, saar2003, desort2007, huelamo2008, boisse2011,  dumusque2011-03, lovis2011, jeffers2013, cabot2021, roettenbacher2021}.  These various types of photospheric velocities imprint on a star's spectrum in different, potentially quasi-periodic ways and evolve on a range of timescales\footnote{Note that there exist other potential sources of photospheric velocities we do not discuss in as much detail (e.g.,evershed flows, moat flows, plage inflows, meridional flows, flares, variable gravitational redshift, etc.), some of which arise from the sources detailed above.}.

Pressure gradients moving through the convective zones of stars result in p-mode oscillations with a timescale of a few minutes, where the frequency and amplitude of these oscillations increases with \teff\ and as stars evolve off the main sequence \citep{mayor2003, bouchy2005, kjeldsen2005, arentoft2008}.  This movement can cause RV variations from 10 \cms\ up to approximately 1 \ms\ for main-sequence stars \citep{dumusque2011-11, chaplin2019}.

Solar-type stars will also exhibit granulation patterns, which arise from convection in the outer layers of the star \citep{dravins1982, lanza2019-03, kjeldsen1995, lindegren2003, nordlund2009, dumusque2011-01, cegla2018, cegla2019}.  Upflows in the middle of granulation cells appear blueshifted while the downflows in the narrow, dimmer edge regions appear redshifted.  This uneven balance between the upflow and downflow regions creates a net RV blueshift, known as convective blueshift, which can lead to asymmetries in spectral lines.

The granulation pattern changes on the timescale of a few minutes to hours, which integrates to different net RV shifts across the surface of the star.  These changes result in varying magnitudes of the convective blueshift and therefore likewise vary the resultant spectral line-shape changes.  This effect can introduce random RV variations of 0.4 to 0.8 \ms, an effect that increases with the \teff\ of the star \citep{meunier2015}.

Supergranulation describes large cells outlined by the magnetic network; it has only been measured on the Sun where cells can persist for hours to up to two days \citep{rieutord2010, rincon2018, meunier2019}.  Changes in supergranulation cells give rise to similar issues as granulation and can introduce RV variations of 0.3 to 0.7 \ms\ \citep{meunier2015}.

Strong magnetic fields can also generate activity features, i.e., darker starspots or brighter faculae in the photosphere and bright plages in the chromosphere \citep{saar1997, hatzes2002, saar2003, desort2007, huelamo2008, boisse2011, dumusque2011-03, lovis2011, jeffers2013}.  This magnetic activity will suppress convection in a star and change the magnitude of the convective blueshift relative to a quiet photosphere.  With Solar data, the effect of this magnetic activity was measured to result in a net redshifted RV change of 0.4 to 1.4 \ms integrated over the surface of the Sun \citep[][]{meunier2010}.  The expected RV variation due to magnetic activity features will change for each star depending on the type of star and the nature of the activity feature.

Activity features rotate in and out of view as the  star rotates.  Spots, which have a lower temperature than the rest of the star, suppress flux while faculae and plages, which instead have a higher temperature, increase the flux in that region.  The presence of activity features therefore changes the integrated flux distribution of the star.  As a star rotates, the side of the star rotating towards the observatory appears blueshifted while the side rotating away appears redshifted.  If the same amount of flux is coming from both sides (i.e., the star is featureless), these effects cancel each other out.  Changes in flux due to activity features can break that balance and introduce up to 10-100 \ms\ variations depending on the specific properties of the star, such as its $v \sin i$, and the properties of the activity features, such as their size, number, and contrast \citep{saar1997, meunier2010}.  The different temperatures of activity features locally modify absorption and emission processes and produce asymmetry in the integrated spectral line profiles that vary with stellar rotation.

Traditionally, stellar signals have been decorrelated from radial-velocity measurements with the use of ``activity indicators.''  These indicators aim to gauge the level of magnetic activity on the target star and/or specifically the presence of activity features for each exposure so that their effects can be removed from RV time series \citep[e.g.,][]{boisse2009, dumusque2011-11, figueira2013, holzer2021}.  Magnetic activity on the star has been shown to correlate with localized spectral features including emission in the core of Ca II H\&K lines \citep[396.96 nm and 393.47 nm respectively;][]{saar1998, meunier2013}, the Ca infrared triplet \citep[849.8, 854.2, and 866,2 nm;][]{saar2000}, and the H-$\alpha$ line \citep[656.28 nm;][]{skelly2008, robertson2014, giguere2016}.

Other popular indicators include properties of the cross-correlation function (CCF) commonly used to derive RVs.  These include various CCF bisector asymmetry measurements \citep[e.g.,][]{queloz2001, povich2001} or the full-width half max of the CCF \citep[e.g.,][]{queloz2009}.  The CCF can be thought of as an average of all line shapes in the spectrum, and is therefore only sensitive to line-shape changes that appear in most lines.  This averaging means RVs derived from the CCF can only be swayed by line asymmetries that persist in the derived CCF.  Methods that disentangle stellar signals by modeling asymmetries in the CCF will likewise only know about the most common line-shape changes as smaller or more unique changes are likely to be averaged out.

Linearly decorrelating RVs against classic activity indicators has not been successful at disentangling stellar signals to sub-meter-per-second precision \citep{fischer2016}.  Recently, more advanced methods have been proposed for deriving activity indicators \citep[e.g.][]{haywood2020} and for disentangling stellar signals from true center-of-mass RV shifts.  Gaussian process (GP) models have been used to more flexibly model stellar signals \citep{haywood2014, rajpaul2015, faria2016, rajpaul2017, angus2018, jones2021, gilbertson2021}.  Methods using different activity indicators and a Bayesian framework were found to more efficiently recover planets in the face of red noise from stellar signals \citep{dumusque2017}.

There has also been a move towards capturing the effects of stellar activity at the level of the 1D spectrum, i.e., before calculating the CCF and extracting RVs \citep[e.g.,][]{davis2017, thompson2017, meunier2018, dumusque2018, wise2018, holzer2021, jones2021}.  The use of pixel-level statistical techniques has revealed that different lines show different behaviors and levels of sensitivity to stellar activity.

With many promising methods being developed to address the issue of stellar signals, we present here a head-to-head comparison of many of these methods on real data.  For four stars---HD~101501, HD~26865, HD~10700, and HD~34411---the \name\ released high-fidelity Extreme Precision Spectrograph (\expres) data that are representative of next-generation spectrographs as well as differential photometry from the Fairborn Automatic Photoelectric Telescopes (\apt)  \citep{zhao2020-rn}.  Eleven teams tested 22 different methods\footnote{This includes 15 unique methods and their variations.} on the data provided.  All methods use the data products provided (i.e., spectra, CCFs, RVs, and/or derived activity indicators), which allows us to compare the performance of methods on exactly the same data.  Because the comparison is done using real data, we do not know exactly the nature of the stellar signals nor what planetary signals may exist within each data set.  Comparison of method results are therefore done relative to other methods.

The data and targets are described in Section \ref{sec:data}.  Section \ref{sec:methods} gives an overview of all methods tested and highlights commonalities between methods (with longer method descriptions included in the Appendix).  The resulting RVs from the different methods are compared in Section \ref{sec:results}.  Section \ref{sec:summary} gives a summary of all methods and the pertinent results.  Section \ref{sec:discussion} discusses the different assumptions made by methods that define the current state of the field.  From there, we make suggestions for future method development and data challenges.  We conclude in Section \ref{sec:conclusion}.

\section{Data}\label{sec:data}

\begin{deluxetable*}{l | c c | c c | c c | c c}[bht!]
\tabletypesize{\scriptsize}
\tablecaption{Stellar Parameters \label{tab:stars}}
\tablehead{
 \colhead{} & \multicolumn{2}{c|}{HD 101501} & \multicolumn{2}{c|}{HD 26965} & \multicolumn{2}{c|}{HD 10700} & \multicolumn{2}{c}{HD 34411}
}
\startdata 
Spectral Type & G8V &  & K1V &  & G8V &  & G0V &  \\ 
$V$ & 5.34 & (d) & 4.43 & (d) & 3.50 & (d) & 4.71 & (d) \\ 
$B$-$V$ & 0.74 & (d) & 0.82 & (d) & 0.72 & (d) & 0.62 & (d) \\ 
\rhk & -4.483 $\pm$ -0.002 & (f) & -4.928 $\pm$ -0.002 & (f) & -4.976 $\pm$ -0.002 & (f) & -5.085 $\pm$ -0.002 & (f) \\ 
Dist. [pc] & 9.541 $\pm$ 0.012 & (e) & 4.98 $\pm$ 0.006 & (e) & 3.65 $\pm$ 0.002 & (i) & 12.484 $\pm$ 0.034 & (e) \\ 
RV [km s\textsuperscript{-1}] & -5.6 $\pm$ 0.08 & (e) & -42.269 $\pm$ 0.0002 & (e) & -16.597 $\pm$ 0.0002 & (e) & 66.57 $\pm$ 0.08 & (g) \\ 
$L_\mathrm{star}$ [$L_{\odot}$] & 0.609 $\pm$ 0.009 & (b) & 0.457 $\pm$ 0.002 & (e) & 0.52 $\pm$ 0.03 & (h) & 1.732 $\pm$ 0.022 & (b) \\ 
$R_\mathrm{star}$ [$R_{\odot}$] & 0.86 $\pm$ 0.02 & (c) & 0.83 $\pm$ 0.02 & (c) & 0.82 $\pm$ 0.02 & (c) & 1.28 $\pm$ 0.04 & (c) \\ 
$M_\mathrm{star}$ [$M_{\odot}$] & 0.9 $\pm$ 0.12 & (c) & 0.8 $\pm$ 0.11 & (c) & 0.99 $\pm$ 0.13 & (c) & 1.08 $\pm$ 0.14 & (c) \\ 
$T_\mathrm{eff}$ [K] & 5502 $\pm$ 25 & (c) & 5092 $\pm$ 25 & (c) & 5333 $\pm$ 25 & (c) & 5873 $\pm$ 25 & (c) \\ 
log $g$ & 4.52 $\pm$ 0.028 & (c) & 4.51 $\pm$ 0.028 & (c) & 4.6 $\pm$ 0.028 & (c) & 4.26  $\pm$ 0.028 & (c) \\ 
$[$Fe/H$]$ & -0.04 $\pm$ 0.01 & (c) & -0.3 $\pm$ 0.01 & (c) & -0.53 $\pm$ 0.01 & (c) & 0.1 $\pm$ 0.01 & (c) \\ 
Age [Gyr] & 3.5$\substack{+2.8 \\ -2.2}$ & (c) & 12.8$\substack{+1.6 \\ -2.9}$ & (c) & 12.4$\substack{+1.8 \\ -3.1}$ & (c) & 4.8$\substack{+1.0 \\ -0.8}$ & (c) \\ 
$v \sin i$ [km s\textsuperscript{-1}] & 2.2 $\pm$ 0.7 & (c) & 0.5 $\pm$ 0.7 & (c) & 1.6 $\pm$ 0.7 & (c) & 0.1 $\pm$ 0.7 & (c) \\ 
$P_\mathrm{rot}$ [days] & 17.1 & (a) & 40 & (a) & 34 & (a) &  &  \\ 
\enddata 
\tablecomments{(a) \cite{baliunas1996}; (b) \cite{boyajian2012}; (c) \cite{brewer2016}; (d) \cite{ducato2002}; (e) \cite{gaiadr2}; (f) \cite{isaacson2010}; (g) \cite{nidever2002}; (h) \cite{pijpers2003}; (i) \cite{vanleeuwen2007}}
\end{deluxetable*}

\begin{figure*}[thb]
\centering
\includegraphics[width=.65\textwidth]{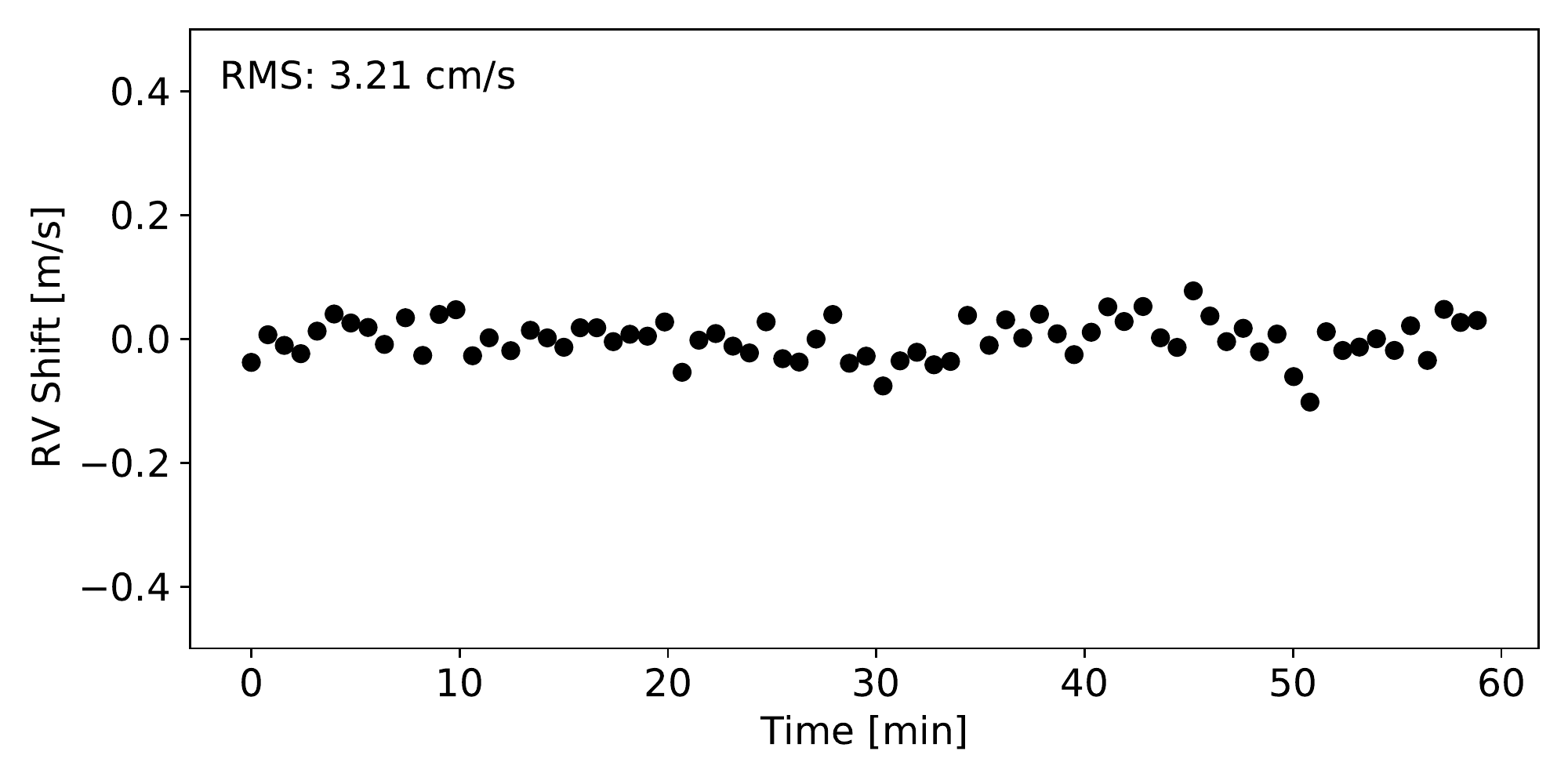}
\caption{Perceived shift in LFC spectra in units of \ms\ across an hour of consecutive LFC exposures with a linear trend removed.  These perceived shifts are attributed to variations in the instrument and therefore give a measure of how stable the instrument itself is.  The RMS of shifts across this hour is given in the top-left corner.}
\label{fig:stableLfc}
\end{figure*}

The data sets for the \essp\ include spectroscopic data from \expres\ and ground-based photometric measurements from the \apt\ for four targets---HD~101501 (61 UMa), HD~26965 (40 Eri), HD~10700 (\tcet), and HD~34411 ($\lambda$ Aur).  Here, we describe the \expres\ and \apt\ instruments, as well as the four targets.  We provide benchmarks for the amount of RV scatter that is expected for the \expres\ instrument and pipeline in the case where there are minimal contributions from stellar signals.  Stellar parameters for each target are tabulated in Table \ref{tab:stars}.

\subsection{Spectroscopic Data From \expres}\label{sec:expresData}
\expres\ is an optical ($390-780$ nm), fiber-fed spectrograph with a median resolution of $R\sim137,000$ \citep{jurgenson2016}. The instrument was fully commissioned at the 4.3-m Lowell Discovery Telescope (\ldt) \citep{levine2012} near Flagstaff, AZ in January 2019 and is being used for a RV planet survey on about 125 (partial) nights per year. The spectrograph is housed in a vacuum enclosure to achieve temperature and pressure stabilization. A Menlo Systems laser frequency comb \citep[LFC;][]{wilken2012, molaro2013, probst2014, probst2020, milakovich2020} ranging from $\sim$490-730 nm is used for precise wavelength calibration.

The instrument calibration stability for \expres\ ranges between 3-7 \cms\ as measured by consecutive LFC spectra taken over thirty minutes to an hour \citep{blackman2020}.  Figure \ref{fig:stableLfc} shows the RV scatter over an hour of consecutive LFC exposures.  The RMS is 3.21 \cms\ after a linear trend is removed.  The linear trend is thought to be due to changing instrument temperature and is accounted for in precision RV work via the wavelength calibration.  The instrument calibration stability can be thought of as the minimum RMS achievable by the \expres\ hardware as it measures the degree of scatter that cannot be calibrated out.

An exposure meter picks off 2\% of the light from behind the fiber entrance to the spectrograph to monitor the photon flux for chromatic barycentric corrections. This exposure meter system also terminates exposures when the target signal-to-noise ratio (SNR) of 250 per pixel at a wavelength of about 550 nm is reached. 

Two or three consecutive exposures, separated only by read-out time, are obtained for each target star per night to improve the nightly-binned precision \citep{brewer2021}. The on-sky, analytical single-measurement precision for exposures with a SNR of 250 (per pixel at $\lambda$=550 nm) is about 0.3 \ms\ \citep{petersburg2020}.  This matches the typical intranight RMS scatter for consecutive observations.

One-dimensional spectra are extracted using a flat-relative, optimal extraction pipeline \citep{zechmeister2014, petersburg2020}.  Extracted spectra were made available to the \essp\ participants along with chromatic barycentric-corrected wavelengths \citep{blackman2017}.  Two sets of wavelengths are provided: one set with a classic polynomial wavelength solution, and one set generated using \excalibur, a hierarchical, non-parametric wavelength solution \citep{zhao2021}.  The provided spectra also include a model of telluric lines generated using \selenite\ \citep{leet2019}, a continuum model, and the associated combined flat image that can be used to recover photon counts\footnote{This is needed since the spectra are extracted relative to this flat image.}.

In addition to extracted spectra, forward-modeled RVs, cross-correlation functions (CCF), and classic activity indicators were provided for each observation.  Theses are described in more detail in the following subsections.  All teams used the provided spectra, CCFs, RVs, and activity indicators as inputs to their methods, thereby ensuring a consistent comparison between the different method results.  Table \ref{tab:targ_rvs} gives the number of RV measurements, the number of nights on which spectra were acquired, and the time baseline for each data set.

\subsubsection{Default RVs}\label{sec:default_rvs}
The standard \expres\ pipeline derives RV measurements using a forward-modeling, chunk-by-chunk technique  \citep{petersburg2020}.  We found the chunk-by-chunk (CBC) RVs to have consistently lower RV scatter than the CCF RVs, and so methods were asked to use the CBC RVs as the default RVs.  A template spectrum is constructed using three consecutive observations of a given target star taken on one night.  Each observed spectra is then broken into two-angstrom chunks that are shifted and scaled to match this template spectrum.  The more chunks there are, the more independent measurements can be derived for the RV; two-angstrom chunks optimize having many chunks while still ensuring each chunk has at least one spectral line.

CBC RVs are derived for each exposure by finding the weighted average of all chunks in a spectra.  The weights for this average are empirically generated based on the stability of each chunk across all observations.  Chunks that are more stable over time are weighted higher while chunks that return higher RV scatter are down weighted.  This reduces the contribution from chunks swayed by, for example, telluric lines, stellar variability, etc.\ and chunks with no spectral lines or containing little RV shift information.  CBC RVs for all four targets are given in Table \ref{tab:cbc_rvs}.

\begin{figure*}[htb]
\centering
\includegraphics[width=.85\textwidth]{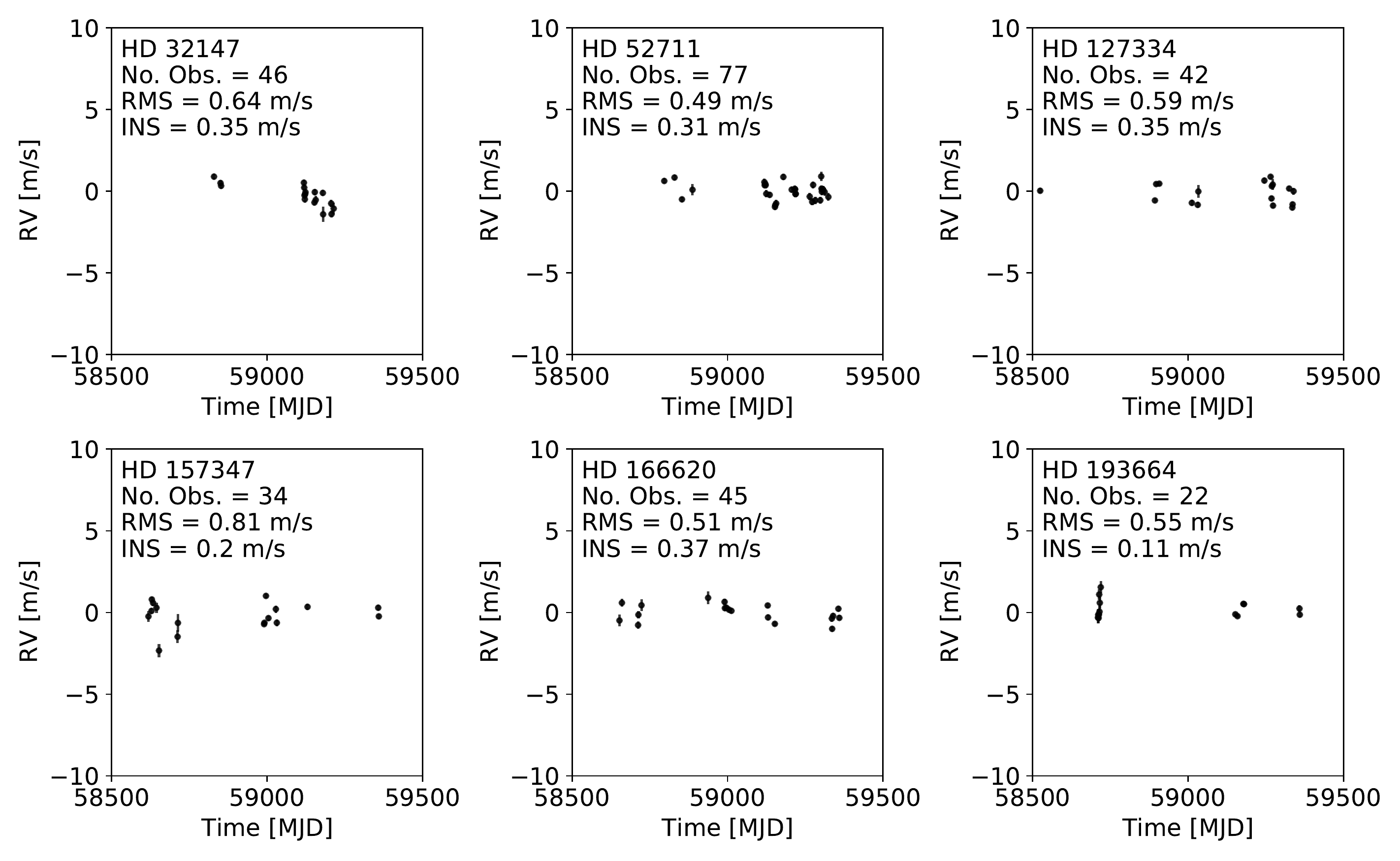}
\caption{\expres\ RVs for six quiet stars.  Shown RVs are derived using a chunk-by-chunk (CBC) forward modeling scheme and binned by night.  The RMS of these nightly-binned RVs are given in the top-right corner along with the average intra-night scatter (INS).}
\label{fig:stableStars}
\end{figure*}

CBC RVs derived from on-sky \expres\ data of chromospherically quiet stars return sub-meter-per-second RMS and intra-night scatter (INS) that matches the analytical 0.3 \ms errors.  Figure \ref{fig:stableStars} depicts RVs from six photospherically quiet stars, which are not part of this study, observed with \expres.  The nightly-binned RV RMS of these pipeline CBC RV measurements range from 0.5 to 0.8 \ms.  The average INS over nights (using only nights with more than one observation) ranges from 0.1 to 0.4 \ms.  These stars demonstrate the RV precision achievable by \expres\ data in the absence of strong photospheric velocities adding scatter.  Complete mitigation of RV contribution from stellar signals should result in a similar final RMS value.

\begin{deluxetable}{l c c l}[tbh]
\tabletypesize{\scriptsize}
\tablecaption{Spectroscopic Observations \label{tab:targ_rvs}}
\tablehead{
\colhead{Target} & \colhead{No. Obs.} & \colhead{Nights} & \multicolumn{1}{c}{Date Range}}
\startdata
HD~101501 & 45 & 22 & Feb. 10, 2019 - Nov. 26, 2020 \\
HD~26965 & 114  & 37 & Aug. 20, 2019 - Nov. 27, 2020 \\
HD~10700  & 174 & 34 & Aug. 15, 2019 - Nov. 27, 2020 \\
HD~34411  & 188 & 58 & Oct. 8, 2019 - Nov. 27, 2020 \\
\enddata
\end{deluxetable}

\begin{deluxetable}{lccc}[tbh]
\tabletypesize{\scriptsize}
\tablecaption{Chunk-by-Chunk RVs \label{tab:cbc_rvs}}
\tablehead{
\colhead{Target} & \colhead{Time [MJD]} & \colhead{RV [\ms]} & \colhead{Err. [\ms]}
}
\startdata
HD 101501 & 58524.466 & -0.338 & 0.322 \\ 
HD 101501 & 58524.491 & 2.38 & 0.325 \\ 
HD 101501 & 58524.497 & 2.66 & 0.308 \\ 
 & \vdots & \vdots & \vdots \\ 
HD 26965 & 58715.487 & -0.101 & 0.435 \\ 
HD 26965 & 58719.469 & -1.85 & 0.368 \\ 
HD 26965 & 58719.472 & -1.44 & 0.408 \\ 
 & \vdots & \vdots & \vdots \\ 
HD 10700 & 58710.457 & 0.075 & 0.388 \\ 
HD 10700 & 58710.458 & -2.25 & 0.377 \\ 
HD 10700 & 58710.46 & -3.03 & 0.387 \\ 
 & \vdots & \vdots & \vdots \\ 
HD 34411 & 58764.475 & 3.47 & 0.324 \\ 
HD 34411 & 58764.477 & 1.98 & 0.34 \\ 
HD 34411 & 58764.479 & 4.8 & 0.314 \\ 
 & \vdots & \vdots & \vdots \\ 

\enddata
\tablecomments{A stub of this table is provided here for reference; the full RV data sets are published online.}
\end{deluxetable}

\subsubsection{Default CCFs}\label{sec:default_ccfs}
The \essp\ provided CCFs as well as the resultant CCF RVs for each spectra.  These CCFs were generated using the code described in \citet{fordEchelleCCFs}.  They make use of CCF masks based on the publicly available \espresso\ masks of the closest matching spectral type with a rectangular window function.  RVs are derived from the CCFs by fitting each CCF to an inverted Gaussian and taking the mean of this Gaussian to be the CCF RV.  Due to differences in the weighting schemes between the CBC pipeline and construction of a CCF, it is expected that the two methods carry different sensitivities to changes in the spectra.

Since the \expres\ pipeline returns flat-relative extractions, it was important to account for the varying SNR of each line.  Lines for the CCF were weighted using the product of the \espresso-mask-provided weight and a constructed weight factor based on the median signal-to-noise (SNR) ratio (assuming only photon-noise).  For lines that show up in multiple orders, the SNR weight factor was computed separately for the line in each order.

Lines that overlap with a telluric feature (as identified by \selenite) during any observation were rejected for all observations.  Lines that were shifted beyond the edge of a given order during any observation were excluded from use within that order for all observations.  Only pixels with a wavelength calibration from the LFC ($\sim$490-730 nm) were used to construct the CCFs.

\subsubsection{Default Activity Indicators}\label{sec:default_indicators}
Each observation was accompanied by several common activity indicators and their empirically determined errors.  The spectroscopic activity indicators provided were the $S$-value---a measure of the emission in the core of the Ca II H\&K lines \citep{meunier2013, saar1998}---and measures of changes in the  \ha\ line core emission \citep{skelly2008, robertson2014, giguere2016}.  Both the \ha\ emission---a measure of the depth of the normalized \ha\ line---and the equivalent width of the \ha\ line were provided.

The provided activity indicators also include a number of indicators derived from the CCF.  The difference in the center of the CCF (whether measured as the bisector of the CCF or the mean of a Gaussian fit) at the top of the CCF as compared to the bottom of the CCF is given as the skew in the CCF bisector (i.e., the CCF bisector inverse span or CCF BIS) \citep{queloz2001} or the velocity span indicator (i.e., \vspan) \citep{boisse2011} respectively.  The top/bottom of the CCF is determined as either a percentage of the total depth of the CCF or defined as a certain sigma away where sigma is the spread of a Gaussian fit to the CCF (for the CCF BIS and \vspan\ respectively).  Varying spectral line profiles will widen the CCF and result in changes to the CCF full width at half maximum (FWHM), which is often used as an activity indicator \citep[e.g.,][]{queloz2009}.  We also provide the results of fitting the CCF to various, asymmetric profiles, such as a bi-Gaussian \citep{figueira2013} or a skew normal probability density function \citep{simola2019}, where the asymmetry parameter of these profiles can serve as an activity indicator.

Analytical errors are provided where possible; otherwise, empirical errors were determined by finding the spread in calculated indicators for nine chromospherically quiet stars\footnote{HD~32923, HD~34411, HD~84737, HD~86728, HD~158633, HD~166620, HD~182488, HD~186427, HD~217014}.  Using a total of approximately 400 observations of these seven stars, a histogram of indicator values was plotted to reveal a Gaussian shape.  The standard deviation of a Gaussian fit to this histogram is taken to be the empirical error for the given activity indicator\footnote{More specifics about how indicators were derived and each indicator's associated empirical errors can be found at \url{http://exoplanets.astro.yale.edu/science/activity.php}.}.

\subsection{Photometry from the \apt}\label{sec:aptData}
Ground-based photometry for all four \essp\ target stars was obtained with either the T4 0.75-m or T12 0.8-m Automatic Photoelectric Telescope (\acronym{APT}) at Fairborn Observatory in southern Arizona.  T4 observed HD~101501, HD~26965, and HD~10700, while T12 observed HD~34411.  Table \ref{tab:targ_phots} describes the number of photometric observations for each target.

\begin{figure*}[t]
\centering
\includegraphics[width=.90\textwidth]{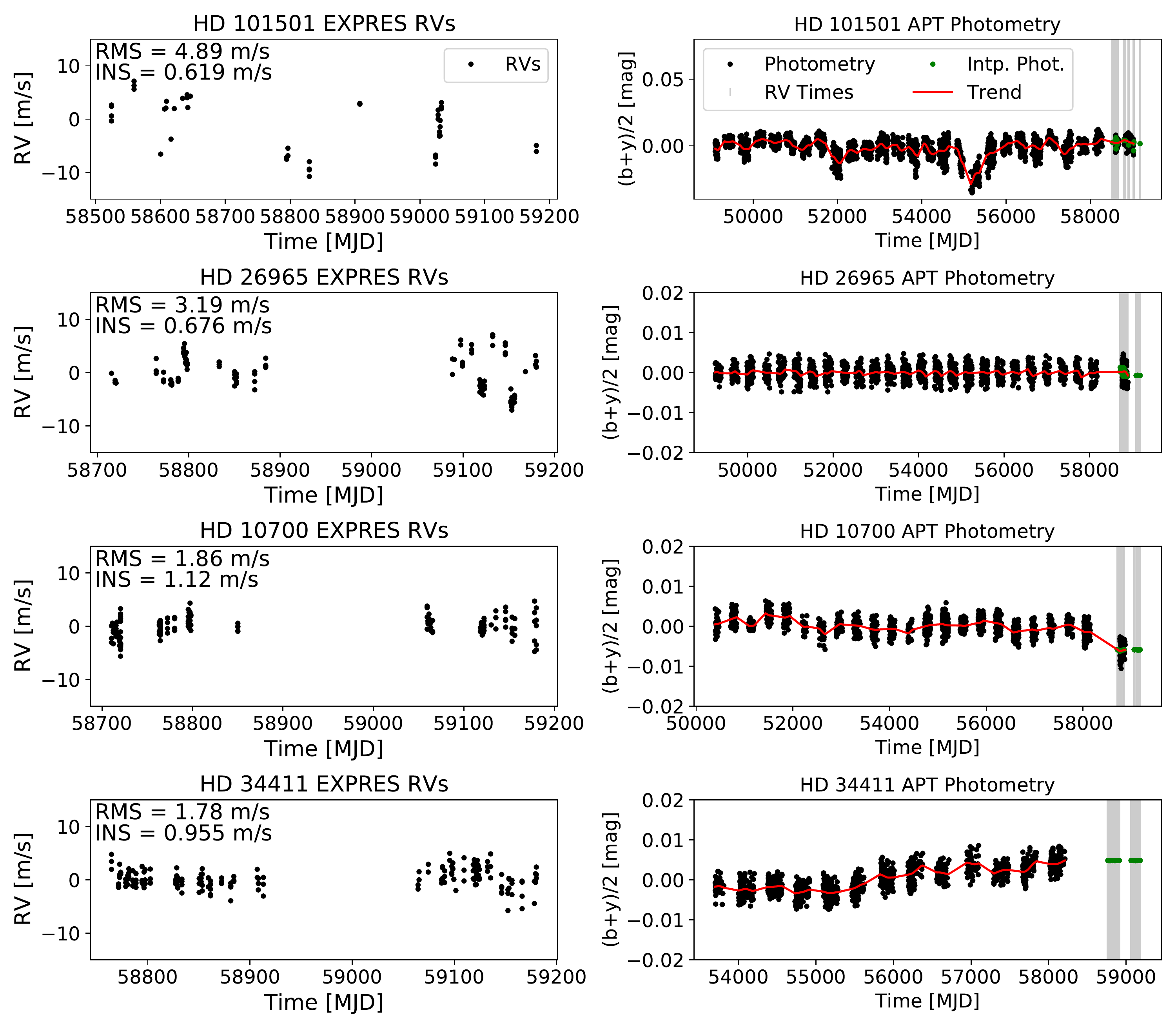}  
\caption{(Left column) \expres\ radial velocities obtained in 2019 and 2020.  (right column) Several years of APT ground-based differential photometry obtained for the four \essp\ stars. Both RVs and photometry time series are plotted with the median value subtracted.  The photometric smoothing model constructed using a 100-day window for each time series is overplotted as a red curve over the photometry (where the photometry is shown before any detrending).  The time stamps of the \expres\ RVs are marked in the right column by gray, vertical lines. The RVs were not taken simultaneously with the photometry with most of the RVs taken after the last photometric observation for all targets but HD~101501.  The GP interpolation/extrapolation of the photometric data to the RV timestamps are over plotted as green points.}
\label{fig:rvNphts}
\end{figure*}

The T4 APT is equipped with a single channel photometer that uses an EMI 9124QB bi-alkali photomultiplier tube to measure the difference in brightness between the program star and three nearby comparison stars in the Str\"{o}mgren $b$ and $y$ passbands.  The T12 APT has a two-channel photometer that uses a dichroic filter to separate the Str\"{o}mgren $b$ and $y$ passbands allowing separate EMI~9124QB photomultiplier tubes to measure the two colors simultaneously.  To improve the photometric precision, we combine the differential $b$ and $y$ magnitudes into a single $(b+y)/2$ ``passband".  The right hand column of Figure~\ref{fig:rvNphts} shows the light curves of each star, spanning between 13 and 28 observing seasons.  

The precision of a single observation taken with the \apt, as measured from pairs of constant comparison stars, is around 0.0010 to 0.0015~mag on good nights. The T4 and T8 (a twin of T12) APTs are described in \cite{henry1999}, where further details of the telescope, precision photometers, observing, and data reduction procedures can be found.

In each photometric data set, we identify a long-term trend that is modeled by applying Gaussian smoothing to the light curve with a 100-day window.  A window of 100 days was chosen to find trends on the order of observing seasons while preserving signals that occur on the timescale of individual stellar rotations.  Using a long, 100-day window returns a coarse trend that is insensitive to short-term variations since they will be averaged over within the window.  Given that the rotation rates of the target stars, which range from 17-34 days where known, are all well under 100 days, the coarse trend preserves signals on the timescale of the stellar rotation period.  Figure~\ref{fig:rvNphts} plots the photometry as black points without any detrending.  The 100-day window, coarse trend is overplotted as a red curve.  The structure that can be seen in the red trend curve is best understood as variations from different observing seasons, potentially due to overall brightness changes of the star (e.g., activity cycles).  Table \ref{tab:phot} summarizes the photometric measurements and  this smooth trend for all four targets.

The photometric data were interpolated to the time stamps of the given spectroscopic data and associated RVs using a Gaussian process (GP) model with a quasi-periodic kernel \citep{rasmussen2006} and implemented with the \george\ package \citep{george}. This kernel depends on four hyperparameters, $\phi = \{a^2, \lambda_e, \lambda_p, P_{\rm GP}\}$, corresponding to the covariance amplitude, a decay timescale (which is related to the typical spot evolution timescale), a smoothing parameter for the periodic term, and a periodic timescale (which is related to the stellar rotation period), respectively. This kernel is used frequently for photometric modeling and stellar activity characterization in the literature \citep[e.g.][]{haywood2014, angus2018}.

\begin{deluxetable}{l c c l}[tb]
\tabletypesize{\scriptsize}
\tablecaption{Photometric Observations \label{tab:targ_phots}}
\tablehead{
\colhead{Target} & \colhead{No. Obs.} & \colhead{Nights} & \multicolumn{1}{c}{Date Range} }
\startdata
HD~101501 & 3290 & 2113 & Apr. 18, 1993 - Jun. 22, 2020 \\
HD~26965 & 1631 & 1500 & Sep. 9, 1993 - Feb. 20, 2020 \\
HD~10700 & 1369 & 1007 & Nov. 5, 1996 - Jan. 24, 2020 \\
HD~34411 & 1214 & 816 & Nov. 25, 2005 - Apr. 3, 2018 \\
\enddata
\end{deluxetable}

\begin{deluxetable}{l c c c}[tb]
\tabletypesize{\scriptsize}
\tablecaption{Photometry and Long-Term Trend \label{tab:phot}}
\tablehead{
\colhead{Target} & \colhead{Time [MJD]} & \colhead{$(b+y)/2$ [mag]} & \colhead{Trend [mag]}
}
\startdata
HD 101501 & 49095.696 & -0.00145 & -0.653 \\ 
HD 101501 & 49095.782 & -0.0023 & -0.653 \\ 
HD 101501 & 49096.783 & 0.00425 & -0.653 \\ 
 & \vdots & \vdots & \vdots \\ 
HD 26965 & 49239.941 & -0.00231 & -2.29 \\ 
HD 26965 & 49245.933 & 0.00084 & -2.29 \\ 
HD 26965 & 49246.93 & 0.00139 & -2.29 \\ 
 & \vdots & \vdots & \vdots \\ 
HD 10700 & 50392.762 & -0.00435 & -2.63 \\ 
HD 10700 & 50396.743 & 0.00115 & -2.63 \\ 
HD 10700 & 50397.735 & 0.00325 & -2.63 \\ 
 & \vdots & \vdots & \vdots \\ 
HD 34411 & 53699.829 & 0.00075 & -1.11 \\ 
HD 34411 & 53700.842 & 0.00265 & -1.11 \\ 
HD 34411 & 53702.821 & -0.00085 & -1.11 \\ 
 & \vdots & \vdots & \vdots \\ 

\enddata
\tablecomments{A stub of this table is provided here for reference; the full photometric data sets are published online.}
\end{deluxetable}

A GP was trained on the most recent six years of APT data for each star after first determining the best-fit hyperparameters via nested sampling.  While the GP regression returned reasonable interpolated median values and 1$\sigma$ uncertainties, it failed to estimate well-principled extrapolated photometric values for RV timestamps falling after the last photometric measurement. This behavior is expected past a few times the typical spot lifetimes on stars, where the lifetime of a spot is typically of order tens of days, but may be longer for young stars \citep{bradshaw2014,giles2017}.

\subsection{Targets}
The four \essp\ stars, as described in Table \ref{tab:stars}, are being observed as part of the \expres\ 100 Earths survey \citep{brewer2021}.  The targets range in activity level, as predicted by \rhk\ values.  Figure \ref{fig:rvNphts} shows the measured CBC RVs and photometry.

HD~101501 is a G8V star and is the most chromospherically active \cite[\rhk$=-4.483$,][]{isaacson2010} of the four \essp\ targets. The \expres\ RVs exhibit an RMS scatter of 4.89 \ms. A GP model of this data preconditioned on photometry found a statistical preference for an activity-only model \citep{cabot2021}.  The provided HD~101501 data set has the fewest number of observations of the four, though a longer time baseline.

HD~26965 is a K1V star with \rhk$=-4.928$ \citep{isaacson2010} and exhibits an RV RMS scatter of 3.19 \ms.  Previous RV analysis of HD~26965 using \hires\ \citep{vogt1994}, \pfs\ \citep{crane2006}, \chiron\ \citep{tokovinin2013}, and \harps\ \citep{mayor2003} RV data from 2001 to 2016 revealed a periodic signal of about 42.364 days \citep{diaz2018}.  Additional data from the Dharma Planet Survey, which added RVs collected from 2014 to 2015, concluded that there exists a 42.38 day periodic signal from a $K=$ 1.8 \ms planet, and that the stellar rotation rate of the star measured from stellar activity indicators is between 39-44.5 days \citep{ma2018}.  Analysis using the complete set of RV data from the California Legacy Survey (CLS), taken from 1987 through 2020, attributes the periodicity to stellar signals \citep{rosenthal2021}.

HD~10700 (i.e., \tcet) is an older \cite[12.4 Gyr,][]{brewer2016}, G8V star that is chromospherically quiet \cite[\rhk$=-4.976$,][]{isaacson2010}.  The \expres\ RVs exhibit an RMS scatter of 1.8 \ms\ dominated by a five-minute periodic variation that matches what we would expect from p-mode oscillations.  Seven planet candidates have been published, though three of these signals (planet candidates b, e, and d) were later retracted \citep{tuomi2013, feng2017}.  Typically, three to five consecutive  \expres\ observations are taken of \tcet\ per night.  On August 25, 2019 and October 8, 2019, more than twenty back-to-back observations were taken within each night (covering a span of approximately 40 minutes) to achieve a sampling that could resolve p-mode oscillations.

HD~34411 is most similar to the Sun of all four targets; it is a 4.8 Gyr, G0V star \citep{brewer2016, gaiadr2}. The star has low chromospheric activity, with \rhk$=-5.085$ \citep{isaacson2010}.  The \expres\ RVs show an RMS scatter of 1.78 \ms.

\begin{deluxetable*}{l l c c l l}[bht]
\tabletypesize{\scriptsize}
\tablecaption{Teams and Methods \label{tab:methods}}
\tablehead{
\colhead{Team} & \colhead{Method} & \colhead{Input} & \colhead{Run Time} & \colhead{Reference/Contact}
}
\startdata
PennState & \glom & RVs/Indicators & laptop, minutes & \cite{gilbertson2020-02}\\
Sidera & \fdpca & RVs/Indicators & laptop, seconds & Ramirez Delgado et al., in prep\\
 &  &  &  & Dodson-Robinson et al., submitted\\
Porto & \gprn & RVs/Indicators & cluster, hour & Camacho et al., in prep.\\
St.\ Andrews/PennState & \scalpels & CCFs & laptop, seconds & \cite{collier2021}\\
PennState & \scalpels+\glom & CCFs & laptop, minutes & \cite{gilbertson2020-02} \\
OxBridGen & \ccfprime & CCF & desktop, minutes & Baptiste Klein\\
PennState & \fiesta+\glom & CCFs & laptop, minutes & \cite{zhao2022}\\
\mleprv & CCF Linear Regression (LR) & CCFs & laptop, seconds & \cite{debeurs2020}\\
\mleprv & CCF LR + \ha & CCFs/Indicators & laptop, seconds & \cite{debeurs2020}\\
\mleprv & CCF LR + Keplerian & CCFs & laptop, seconds & \cite{debeurs2020}\\
PennState & CCF Mask-VALD & Spectra & laptop, minutes & Alex Wise\\
Warwick & CCF Mask-BIS & Spectra & laptop, minutes & \cite{lafarga2020}\\
Warwick & CCF Mask-RV & Spectra & laptop, minutes & \cite{lafarga2020}\\
Geneva & \lblspec & Spectra & laptop, hours & \cite{dumusque2018}\\
 &  &  &  & Cretignier et al., submitted\\
Geneva & \lblrvel & Spectra & laptop, hours & \cite{cretignier2021}\\
 &  &  &  & Cretignier et al., submitted\\
Geneva & \lblboth & Spectra & laptop, hours & \cite{cretignier2021}\\
 &  &  &  & Cretignier et al., submitted\\
OxBridGen & \pwgp & Spectra & desktop, day & \cite{rajpaul2020}\\
PennState & \dcpca & Spectra & laptop, seconds & \cite{jones2017}\\
PennState & \dcpca+\glom & Spectra & laptop, minutes & \cite{gilbertson2020-02}\\
CCA & \horatio\ Self & Spectra & laptop, hour & Lily Zhao\\
CCA & \horatio & Spectra & laptop, hour & Lily Zhao\\
CCA & \hamlet & Spectra & laptop, minutes & Lily Zhao\\
\enddata
\end{deluxetable*}

\section{Methods}\label{sec:methods}

The \essp\ received submissions from 22 different methods all with the goal of isolating true center-of-mass shifts.  Table \ref{tab:methods} lists all methods along with variations on each method.  The ``Input'' column specifies the primary type of provided data that was input into the method (i.e., the extracted spectra, the CCF, or the CBC RVs along with activity indicators).  The ``Run Time" column gives an estimate of the computational expense of the method by specifying what the method was run on and the order of magnitude of time it took to run.  This, of course, is merely an estimate as the runtime of most methods scale with the number of observations\footnote{Here, most data sets have on the order of 150 observations}.  Related publications for each method are given where available; otherwise, the name of the most pertinent author to contact for each method is listed.

A short description of each method is given below along with any specifics to the implementation represented here and relevant data requirements.  Similar methods are compared and contrasted.  Longer descriptions of each individual method can be found in the Appendix and provided references.

Methods are grouped into subsections according to the type of input data used: RVs with global indicators (\S \ref{sec:rvindMethods}), CCFs (\S \ref{sec:ccfMethods}), or extracted (pixel-level) spectra (\S \ref{sec:lblMethods} and  \S \ref{sec:spectraMethods}).

\subsection{Methods That Use RVs and Classic Activity Indicators as Input} \label{sec:rvindMethods}
Activity indicators aim to gauge the magnetic field strength on the target star, presence of activity features, or otherwise the expected amplitude of stellar signals.  These indicators are global parameters---one value is determined for each spectrum.  One can fit a model relating the activity indicators and apparent RVs in an attempt to remove or mitigate the effect of stellar signals on measured RVs.  Classically, this was done using a simple linear fit.

We present the results of a classic linear decorrelation with the provided activity indicators to serve as a baseline result.  RVs are plotted against the different indicators independently and fit to a line as a function of indicator value.  The fitted line is evaluated at the value of the different indicators, which is then subtracted from the RV measurements.  There is evidence of a phase shift existing between some indicator variation and corresponding RV variation \citep[e.g.,][]{santos2014, collier2019, mortier2021}, which adds scatter to direct comparisons between indicator and RV and limits the efficacy of this linear decorrelation method.

More recent work has developed novel ways of linking indicators to RV measurements and modeling out the chromospheric velocity component of RV measurements \citep{rajpaul2015, barragan2019, gilbertson2020-02, ramirez2021, dodson2021, camacho2021, barragan2022}.  Indicator-dependent methods will only be sensitive to signals that are reflected in the provided indicators; for example, if the used indicators do not track the effects of oscillation or granulation, then these methods will not return models sensitive to these effects.  Teams who used RVs and indicators as input were asked to use the provided forward-modeled CBC RVs and the given indicators.

The Gaussian Process Linear Ordinary Differential Equation (ODE) Maker (\glom), developed by the PennState team, is a Julia package that uses a shared, latent Gaussian Process (GP) to model both RV and indicator time series concurrently.  This makes use of the flexibility of a GP model while also constraining the model with indicator time series to capture only stellar signal related variations.  \glom\ can be thought of as a generalization of the multi-dimensional GP method implemented in \pyaneti\ \citep{rajpaul2015, barragan2019, barragan2022}.  This method requires dense sampling throughout the characteristic timescale of the signal being modeled (e.g., the stellar rotation period for spots).  It is utilized as a part of many of the other submitted methods to the \essp\ that generate an indicator for the presence of stellar signals.  More information can be found in \citet{gilbertson2020-02} or Appendix \ref{sec:glom}.

Fourier Domain Principal Component Analysis (\fdpca), developed by the Sidera team, Fourier transforms RV and indicator time series using nonuniform methods to identify coherent oscillations between multiple series regardless of their relative phases. The Fourier transformed series are decomposed using principal component analysis (PCA) to derive orthogonal axes of variation in the activity indicators and their associated weights.  The results presented here were trained on the RV, \ha\ emission, and  CCF FWHM time series.  The model incorporated increasing numbers of principal components until 95\% of the total variation was captured.  This method requires observations to cover the entire phase range of the signal being modeled.  Observations should be dense in phase space, not just time.  A more in-depth description can be found in Appendix \ref{sec:fdpca}.

The Gaussian Process Regression Network (\gprn) method, developed by the Porto team, models RVs and indicators through a neural net framework where each node is an independent GP model and the weights of each node are also determined by a GP model.  While each node and weight can be represented by an independent GP, hyper-parameters and priors may be shared between models to reduce the number of free parameters.  For the results presented here, one node was defined by a GP with a quasi-periodic kernel while GPs with squared-exponential kernels were used for the weights with no shared hyper-parameters.  The models were trained on the RV and CCF FWHM time series.  The \gprn\ method is still being developed; preliminary results are included here.  A more in-depth description can be found in Appendix \ref{sec:gprn}.

\subsection{Methods That Use the Cross Correlation Function (CCF) as Input} \label{sec:ccfMethods}
The CCF has long been used in endeavors to mitigate the effects of stellar signals.  CCFs are computed by cross correlating a given spectra with a mask tuned to where spectral line centers are expected to appear.  The mask can either be binary (i.e., 1 where there is a line, 0 where not) or incorporate different line widths and window functions.

As this mask is shifted relative to a stellar spectrum, the convolution of the two will give larger or smaller values depending on how well the mask lines up with the spectral absorption lines.  A perfect alignment of the mask with the bottom of every spectral line will result in the lowest cross-correlation value.  The shift that results in the lowest CCF point can then be taken as the RV shift of the spectrum.

In shifting, the CCF will sample the shape of all the spectral lines in the mask, including the wings of these lines.  Lines can be weighted differently according to their depth or their SNR.  The CCF therefore provides a sort of weighted average of all the line shapes in the spectrum.  This makes the CCF a powerful tool when there exists line-shape distortions.  On the one hand, averaging over all lines in a mask can blur out different line-shape changes seen in individual lines; on the other, this averaging will also strengthen the signal of any line-shape changes that are common to many lines and cause the resultant CCF to be asymmetric.


Four methods used the CCF as input.  All four of these methods focus on modeling asymmetries within the CCF, which are attributed to the spectral level line-shape changes known to be caused by stellar signals.  The methods differ in their approach to modeling the CCF shape changes and how these changes are separated from translational shifts.

The Self-Correlation Analysis of Line Profiles for Extraction of Low-Amplitude Shifts (\scalpels) method, submitted by the St.\ Andrews and PennState teams, uses PCA to model the variations in a CCF's auto-correlation function.  Because the auto-correlation function is intrinsically insensitive to translational differences, \scalpels\ is not sensitive to true shifts in the CCFs.  The measured RV time series can then be projected onto the identified principal components to derive and subtract out the shape-driven component of the measured RV while preserving the shift-driven component.  The results presented here use only the first two principal components to guard against incorporating noise into the model.  \scalpels\ operates in the velocity-domain and as such does not require any additional information about the star or dense time sampling.  Using PCA means the model benefits from wider ranges of stellar activity states producing a large range of variation within the CCFs.

\scalpels\ uses PCA in a similar way to \fdpca, where the principal components are used as a new basis with which to construct a denoised measurement of RV shifts due to stellar signals.  \scalpels\ uses PCA on the auto-correlation function of the CCFs while the \fdpca\ method implements PCA on the Fourier series of the RV and indicator time series.  Note that while there is a description of a leave-one-out-framework with \scalpels\ in \cite{collier2021}, no cross-validation framework is implemented for the results submitted here.

The \scalpels+\glom\ method is another use of PCA.  The amplitudes of the first two principal components for each observation, which describe the magnitude of the two largest axes of variation in the CCF auto-correlation function, are treated like activity indicators and input into \glom\ to be co-modeled with the RV measurements.  For the results presented here, the latent GP model used the sum of two \matern\ kernels.  This introduction of a GP model introduces relevant data requirements to the method, such as dense-sampling in time.  More information about both implementations of \scalpels\ can be found in \citet{collier2021} as well as Appendix \ref{sec:scalpels}.

The \ccfprime\ method, submitted by the OxBridGen team, uses higher order derivatives of a GP modeled reference CCF (here a mean combined CCF) to fit shape changes.  While the first GP derivative is sensitive to translational differences, the second derivative and above are instead sensitive to shape changes.  These higher-derivatives are used to recreate the shape-driven component of the measured RVs, which can then be subtracted out.  The \ccfprime\ method is still being developed; preliminary results are included here.  A more in-depth description of the \ccfprime\ method can be found in Appendix \ref{sec:ccfprime}.

The FourIEr Phase SpecTrum Analysis (\fiesta) method, submitted by the PennState team, isolates line-shape changes using a Fourier basis with respect to velocity.  Horizontal, translational differences manifest as a constant shift at all frequencies in this basis.  Shape-driven shifts can therefore be isolated as frequency-dependent shifts.  The results presented here run a PCA on the derived shifts for each frequency and uses the amplitudes from this PCA as input into \glom.  This is similar to how PCA is used within the \scalpels+\glom\ framework (which is distinct from the use of PCA in the \scalpels\ or \fdpca\ methods).  \fiesta\ requires careful normalization of the CCFs for each observation, as vertical translational differences could be mistaken for a shape change.  More information can be found in \citet{zhao2020}, \citet{zhao2022}, and Appendix \ref{sec:fiesta}.

The \scalpels, \ccfprime, and \fiesta\ methods all implement a change of basis to separate out the shape- and shift-driven components of the measured RVs.  While these methods are conceptually similar, they make different assumptions of the appropriate basis and dimensionality of the variations being modeled.  High SNR observations are more necessary with \ccfprime\ (for more accurate GP derivatives) and \fiesta\ (to allow for incorporating higher frequencies) than with \scalpels.  \scalpels, on the other hand, is more dependent on the assumption that the dominant source of variation that gets captured by the PCA are shape-driven changes from stellar variation.

The CCF Linear Regression method, submitted by the \mleprv\ team, uses machine learning to model variations in the residuals of each CCF as compared to a reference CCF (here a median combined CCF).  Differential CCFs are normalized (in terms of amplitude and overall variance) and then sampled at a small-number of locations across the velocity range of the CCFs.  A larger number of observations per target allows for more sampling locations.  For the results presented here, the CCFs were sampled at four to six locations.  For each target star, a linear regression model was used to fit an associated weight parameter for each of the sampled CCF locations.  In this way, the changes in CCF shape are captured to predict the chromospheric contribution to the RV signal.  This method does not use timing information, and so does not care about the sampling of observations, but does benefit from more observations.

For all four targets, a slightly more complicated CCF Linear Regression model was also implemented, that included the \ha\ emission value for each observation with its own fitted weight parameter to help predict variations due to stellar signals.  For HD 26965, which hosts a proposed planet, a third model that incorporates a weighted Keplerian was also implemented\footnote{The same was not done for the \tcet\ data set.}.  These methods are still being actively developed; the results included here are preliminary.  More information can be found in Appendix \ref{sec:ccfLR}.  The work that inspired this method and was implemented on the solar data is described in \cite{debeurs2020}.

All four methods focus on the shape-changes caused by stellar signals.  Shape changes captured by the CCF are either separated out from translational shifts (i.e., \scalpels) or the amplitude of the shape change is measured.  This measured amplitude is either unitless and used as an activity indicator (i.e., \scalpels+\glom, \fiesta) or used to derive the resultant RV shift due to shape changes (i.e., \ccfprime, CCF Linear Regression).  The separated out translation-driven shifts or the residual RVs from subtracting out shape-driven RV measurements are returned as RVs cleaned of stellar signals.  This separation will not be sensitive to the effects of stellar signals that may manifest as translational shifts within the CCF rather than as a shape change.  Should there exist a manifestation of stellar signals that produces only a strict translation shift to the CCF, this effect will be wholly degenerate with measurements of true center-of-mass movement at the level of the CCF.

All methods attempting to model line-shape changes, such as the four described here, will be helped by data with high resolution.  Higher resolution spectra contain more information about the shape of each spectral line and will therefore more faithfully capture the shape deformations being modeled.  This is true whether the shape changes are being modeled as averaged in the CCF or in the spectra itself.

\subsection{Line-by-Line Methods} \label{sec:lblMethods}
The remaining methods take the full, extracted spectra as input.  Several methods, described in this section, use the spectra to determine preferred lines or regions of spectra to use when deriving RV measurements.  Methods that model variation throughout the complete spectra are described in the following section (\S \ref{sec:spectraMethods}).

Three methods focused on carefully picking which lines to include when constructing a CCF.  It has recently become clear that spectral lines will respond in different ways to stellar variation, both in terms of behavior and magnitude of response \citep{davis2017, thompson2017, meunier2017, wise2018, dumusque2018, cretignier2020-01, jones2021}.  Isolating lines that are less swayed by stellar signals or other occluding effects will help in calculating CCFs and RVs that are resilient to these variations and ultimately more representative of true, center-of-mass shifts in the spectra.

The CCF Mask-\vald\ method, submitted by the PennState team, used line center information from the Vienna Atomic Line Database (\vald) to vet lines included in the CCF mask and remove line blends that may otherwise introduce asymmetries to the derived CCF unrelated to stellar signals.  Additionally, for most lines the dominant effect of stellar variability is to alter the depth of the line.  In the case of blended lines, a depth change in either line will affect the velocity measured from that line \citep{dumusque2018} or from a CCF computed using either of the lines.  Designing a CCF mask that uses only well-isolated lines therefore provides a path to measuring RVs that are less sensitive to stellar signals.  The method also optimizes across a range of CCF mask window widths, which tunes how much of a line's profile is averaged into each CCF point--i.e., a narrower CCF mask window width samples each spectral line's shape at higher resolution but will be noisier.  A truncated Gaussian window function was used for all lines.  The optimal cutoffs for distance between line centers and width of mask window were found empirically by minimizing the RMS of the resultant CCF RVs.  More details can be found in Appendix \ref{sec:wiseCcf}.

The CCF Mask-BIS and CCF Mask-RV methods, both submitted by the Warwick team, use correlations with the BIS activity indicator or the provided CCF RVs to vet lines.  RVs for individual lines were found by measuring the shift in each line center for each line across all exposures.  Each line is fit to a Gaussian and the mean of this fit is taken to be the line center.  Lines were excluded if their RVs were found to scatter greater than 10-15 \ms\ or their RVs were found to be strongly correlated with the BIS or CCF RV (i.e., the Pearson correlation coefficient is greater than some cutoff, where the cutoff depends on the specific indicator used and target).  The RVs of the remaining lines are averaged to compute a combined RV for each observation.  More details can be found in \citet{lafarga2020} and Appendix \ref{sec:warwickCcf}.

Note that CCF Mask-RV is not the only method to use the RV as an activity indicator (see, for example, the discussion of the \horatio\ and \hamlet\ methods below).  This use case assumes that all variation in the measured RVs is dominated by stellar signals.  We know that instrument systematics are not the dominant source of error in these data sets, as seen from \expres\ data of quiet stars (see Figure \ref{fig:stableStars}).  While there are no obvious planetary signals, this does not preclude planet signals on the order of or smaller than the stellar signal amplitude adding variation to the RVs.

All three of the above methods fit lines to a Gaussian profile to determine line parameters---such as line center, width, etc.---or change in line parameters.  The provided \selenite\ telluric model was also used in all three cases to remove lines within $\sim$30 \kms\ of a telluric feature.

The Geneva team also implemented a line-by-line (\lbl) RV analysis.  The \lbl\ RVs were derived relative to a master spectrum using post-processed spectra \citep{dumusque2011-01}.  The provided \expres\ spectra were (1) merged (i.e., all echelle orders were combined), (2) continuum normalized using \rassine\ \citep{cretignier2020-02}, and then (3) further cleaned of tellurics and first-order morphological variations using \yarara\ \citep{cretignier2021}.  Lines returning a poor fit to the master spectrum or exhibiting larger scatter than expected from the median RV error were not included in the final combined RV calculation.

PCA was used to de-noise the results at either the spectral level, denoted here as \lblspec, or produce a metric of variation at the line-by-line RV level, \lblrvel.  At the spectral level, the first three components of a weighted PCA are used to recreate a denoised representation of the spectra.  These de-noised spectra are then used to construct a master spectrum and derive \lbl\ RVs.

PCA was also run on the \lbl\ RVs themselves to identify variations across all lines and across all observations.  Rather than denoising, here PCA is instead used to determine the magnitude of variation that is then treated like an activity indicator against which the combined \lbl\ RVs are decorrelated with a multi-linear regression.  \lbl\ RVs that are derived and decorrelated using RV-level PCA are described as the \lblrvel\ method.

Both methods can also be combined, which is here represented by the \lblboth\ method.  Though  both \lblspec\ and \lblrvel\ use PCA, it is important to note that PCA is used on different data products for the two methods and to different ends.  The difference is similar to the difference between how PCA is utilized in the \scalpels\ method versus the \scalpels+\glom\ method.  More details about deriving \lbl\ RVs and the \rassine\ and \yarara\ methods can be found in \citet{dumusque2018, cretignier2020-02, cretignier2021}.  More information about the \lblspec, \lblrvel, and \lblboth\ implementations represented in this report can be found in Appendix \ref{sec:lblrvs}.

The Pairwise Gaussian Process RV Extraction (\pwgp) method, submitted by the OxBridGen team, breaks the spectrum into chunks and uses GPs to model and align pairs of chunks.  Like was described for the \expres\ piepline (\S \ref{sec:default_rvs}), the behavior of each chunk across all observations is used to determine which areas of the spectrum are more or less sensitive to variation from telluric contamination or stellar signals.  In the limit where each chunk contains one line, which the implementation presented here approaches, the \pwgp\ method can be thought of as an approximate line-by-line approach.  A \matern\ kernel is used in the GP that models and aligns each chunk.  Chunks exhibiting unusually high scatter or strong correlation with activity indicators are excluded.  The RV for each observation is then calculated as a weighted average of the remaining chunks, where the RV error for each chunk is determined via a MCMC analysis.  More information can be found in \citet{rajpaul2020} and Appendix \ref{sec:pwgp}.

For all these methods, there exists a trade off.  Increasing the selectivity of lines or chunks to include will better mitigate the effects of stellar signals and other possible causes of line-shape variation.  Using less data, however, will increase the photon noise.  These methods would all benefit from high SNR observations, which decreases the photon noise that must be contended with.  This allows for confident RV estimates from relatively few, very stable lines.

\subsection{Full-Spectrum Methods}\label{sec:spectraMethods}
While the methods described in the previous section treated each line/chunk as independent, the below methods model the entire spectra as a whole.  Of course, in some ways the methods of the previous section do take into account information across the whole spectra, for example when setting cutoffs using all lines or running PCA on all lines. Unlike previously presented methods, though, these ``full-spectrum'' methods generally operate on all spectral pixels.

The Doppler-Constrained Principal Component Analysis (\dcpca) method, submitted by the PennState team, runs PCA on spectra shifted by the maximum-likelihood RV and uses the resultant PCA amplitudes, a measure of the amount of primary variation present in each exposure, as activity indicators.  Though the PCA is run on the spectra, this use case of PCA is more similar to the \lblrvel\ method (or \scalpels+\glom): the amplitude of the variation, not the axes of variation (i.e., the principal components), are the result of interest.  To cut down on the noise that gets input into the PCA, only the spectral regions surrounding lines included in the default \espresso\ masks used are fed into the PCA.  These indicators are then either linearly decorrelated against RVs (\dcpca) or co-modeled with RVs using \glom\ (\dcpca+\glom).  More information can be found in \citet{jones2021} or Appendix \ref{sec:dcpca}.

Generative residual regression and discriminative residual regression (\horatio\ and \hamlet\ respectively), both submitted by the CCA team, use the pixel-level residuals of observed spectra from a template spectrum to regress against different housekeeping data---such as activity indicators---and derive the contribution from the stellar to the measured RV shifts.  Under a generative framework, \horatio\ uses the \ha\ equivalent width and CBC RVs as labels to derive the activity-component of the measured RVs.  \hamlet\ moves in the other direction---the full residuals of each observation are used to inform the appropriate correction to the measured RVs.  The discriminative framework is slightly more agnostic to the labels used, meaning \hamlet\ is less tied to the information available in the used activity indicators than \horatio\ is.  Both methods use a linear, first-order model and residuals to a reference template constructed using \wobble\ \citep{bedell2019}.

Both \horatio\ and \hamlet\ implement a ``cross-validation" (CV) framework.  This guards against over-fitting as the model is constructed without information from the subset of data that the model is then evaluated at.  For \hamlet, each observation is left out one at a time to construct an independent model.  For \horatio, one eighth of the data is left out at a time to speed up the computation time.

For reference, the ``self" test variant for \horatio\ (\horatio\ Self) is included, wherein all data are used to construct the model.  The only difference, then, between \horatio\ and \horatio\ Self is removing the cross-validation framework that is used to prevent over-fitting.  Because seven eights of the data are still used  to construct the model for the cross-validation version of \horatio\ and the validation data is chosen at random across the time baseline, we do not expect the \horatio\ method to be less informed than the \horatio\ Self method that uses all data points; the cross-validation step only ensures the resultant model is general.  The \horatio\ Self method is presented as a more direct comparison to the RMS metrics reported by other methods that did not employ cross-validation when deriving their reported results.

The \horatio\ and \hamlet\ methods are still being developed; preliminary results are included here.  More information about \horatio\ and \hamlet\ can be found in Appendix \ref{sec:horatio} and \ref{sec:hamlet} respectively.

\section{Results}\label{sec:results}
For each method, teams submitted ``clean RVs" representing the measured RV shift of the star cleaned of stellar signals and other modeled variations leaving only true center-of-mass shifts.  Where directly modeled, the RVs due to the modeled out variations, which we will refer to as ``activity RVs,'' were also submitted along with any indicators that the method derived.

We acknowledge that this chosen name of ``activity RVs'' is imperfect.  The variations being traced by different methods may source from stellar activity features, such as spots, faculae, etc., but could also be due to inherent stellar variation, such as pulsations or granulation, or trace other sources of variation in the spectra from, for example, uncorrected tellurics, instrumental changes, etc.

For some methods, the clean and activity RVs represent different components of the model and so do not sum to the original RVs provided.  The clean RVs and provided activity RVs from all methods are plotted in Appendix \ref{sec:rawResults} along with their Lomb-Scargle periodograms \citep{lomb1976, scargle1982,vanderplas2015}.

\subsection{RV RMS of Method Results}\label{sec:results_rms}
Table \ref{tab:results} gives the change in overall and nightly RMS for each method as compared to the RMS of the provided, uncorrected CBC RVs.  The nightly RMS, or intra-night scatter (INS), represents the average scatter over all nights with more than one observation.  Positive $\Delta$RMS values indicate that the method returned a lower RMS than the original.  Negative $\Delta$RMS values means there was more spread in the returned RVs than in the original provided RVs.  Methods are ordered in the same order as described in the Methods section above (\S \ref{sec:methods}).

\begin{figure*}[hbt!]
\centering
\includegraphics[width=.96\textwidth]{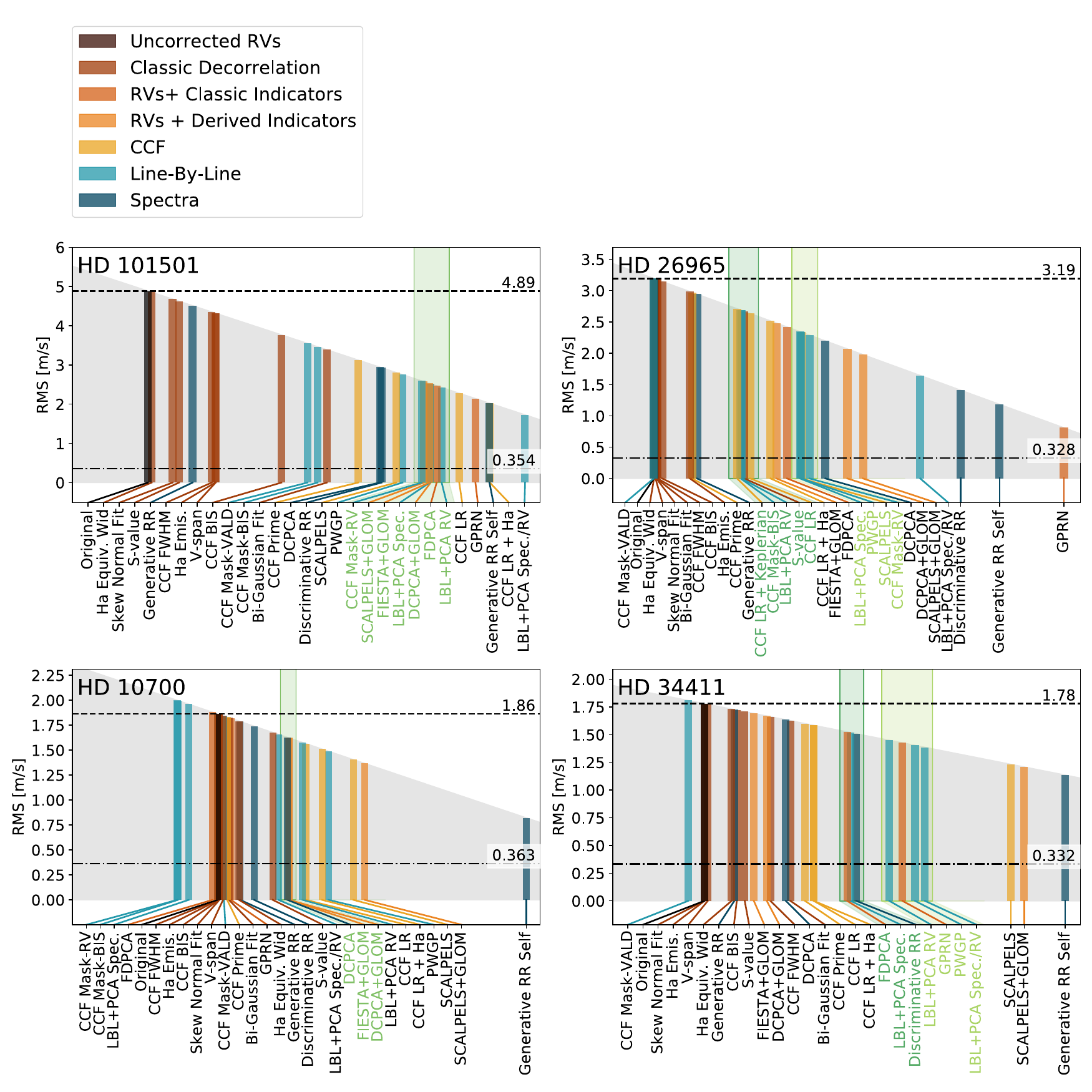}
\caption{Final overall RMS of the clean RVs submitted for each of the four targets.  The height and x-position of each bar scales with the final RMS.  Bars are colored by the type of input data used.  For each target, the RMS of the original, uncorrected \expres\ CBC RVs is shown as a black bar with its height emphasized by a horizontal dashed line across the full plot.  The average intra-night scatter of the \expres\ CBC RVs is also marked with a horizontal dash-dotted line.  Methods returning similar RMS values to each other are emphasized via green shading.}
\label{fig:rmsbar}
\end{figure*}

The final RMS values of the clean RVs for all methods are plotted in Figure \ref{fig:rmsbar}.  The height of each bar as well as its position along the x-axis scales with the overall RMS of the returned clean RVs.  Each bar is mapped to its corresponding method across the x-axis, along which the methods are ordered by decreasing RMS from left to right.  The bars are arranged in this way to emphasize the relative RMS values of the cleaned RVs returned by each method.

Resultant RV RMS gives information about the magnitude of the signal being removed by each model, but does not contain any information about the nature of the signal being removed.  While RMS alone cannot establish the success of a method in mitigating stellar signals, this initial look at the relative final RV RMS of the different methods gives a sense of the amplitude of the signal being removed by each method, which methods are removing a comparable amplitude of signal, and whether any methods are increasing the RV scatter.

Each bar is colored by the type of data the method takes in as input, corresponding to the break down of methods in Section \ref{sec:methods}.  Note that here, all methods that use a sort of activity indicator, classic or newly derived (e.g., amplitudes from PCA, etc.), are grouped together regardless of the input data needed to derive the indicator used.  Other than the methods that decorrelate against a classic activity indicator, methods that take in the same input do not necessarily return similar final RMS values.

The baseline method of decorrelating RVs against classic activity indicators, shown in Figure \ref{fig:rmsbar} as brown bars, does not produce a significant decrease in RMS.  The decrease is modestly significant for HD 101501, the most active of the targets given.

The relative returned RMS of each method differs across the four stars.  Methods that return low RMS for one or some of the targets do not necessarily return low RMS for all of the targets.  The relative RMS of different methods is most different for HD~26965, for which a few methods (\gprn, \hamlet, and \lblboth) return much lower relative RMS values than for the other three targets.  For HD~101501, \lblboth\ and \gprn\ also return among the lowest RMS, as with HD~26965, but with the HD~101501 CCF LR and CCF LR+\ha\ return much lower relative RMS than they do with any other target.  The relative orders of the methods are fairly consistent between HD~10700 and HD~34411.  Recall that the four targets differ in expected activity level, total number of observations, sampling of observations, and number of proposed planet candidates.

For each of the four targets, there are one or more clusters of methods returning a similar RV RMS, which can be seen as overlapping bars in Figure \ref{fig:rmsbar}.  For HD~101501, there is a cluster of methods returning a final RMS of approximately 2.5 \ms, i.e., a 48\% decrease in RMS.  The HD~26965 results exhibits a cluster at 2.7 \ms\ (16\% decrease) and 2.3 \ms\ (28\% decrease).  Note that these RMS values are slightly greater than the 1.8 \ms\ semi-amplitude of the proposed planet \citep{ma2018}.  The HD~10700 (\tcet) results cluster around 1.6 \ms\ (13\% decrease).  The HD~34411 results cluster at 1.5 \ms\ (15\% decrease) and 1.4 \ms\ (22\% decrease).  The methods that are returning similar RMS values and forming these clusters differ in their approach to disentangling stellar signals, and in fact the methods that are clustered together even differ from target to target.

The self test version of \horatio , \horatio\ Self, always returns a lower RMS than the cross-validation implementation of \horatio.  Furthermore, \horatio\ Self is often among the methods returning the lowest RMS value.  \horatio\ and \horatio\ Self only differ in whether there is a safe guard built into the method against over-fitting the model.  The difference in their resultant RMS therefore highlights the difference between an appropriately general model with \horatio\ and a likely over-fitted model with \horatio\ Self. We note that while the \horatio\ class of methods has many free parameters and is therefore particularly vulnerable to over-fitting, some degree of over-fitting may be in play for other methods presented in this paper that did not implement a cross-validation step.  Most methods returned results of a model trained on the same data that they

\begin{rotatetable*}
\begin{deluxetable*}{l | c c | c c | c c | c c}
\tabletypesize{\scriptsize}
\tablecaption{RMS and INS of Cleaned RVs from each Method in Units of \ms \label{tab:results}}
\tablehead{
 \colhead{} & \multicolumn{2}{c|}{HD 101501} & \multicolumn{2}{c|}{HD 26965} & \multicolumn{2}{c|}{HD 10700} & \multicolumn{2}{c}{HD 34411} \\
\hline
 & \colhead{INS } & \multicolumn{1}{c|}{$RMS_{all}$} & \colhead{INS} & \multicolumn{1}{c|}{$RMS_{all}$} & \colhead{INS} & \multicolumn{1}{c|}{$RMS_{all}$} & \colhead{INS} & $RMS_{all}$ \\
\multicolumn{1}{l|}{Original \expres\ CBC RVs} & \colhead{0.62} & \multicolumn{1}{c|}{4.887} & \colhead{0.65} & \multicolumn{1}{c|}{3.195} & \colhead{1.071} & \multicolumn{1}{c|}{1.864} & \colhead{0.944} & \colhead{1.78} \\
\hline
\multicolumn{1}{l|}{Method} & \colhead{$\Delta$ INS} & \multicolumn{1}{c|}{$\Delta RMS_{all}$} & \colhead{$\Delta$ INS} & \multicolumn{1}{c|}{$\Delta RMS_{all}$ } & \colhead{$\Delta$ INS} & \multicolumn{1}{c|}{$\Delta RMS_{all}$} & \colhead{$\Delta$ INS} & \colhead{$\Delta RMS_{all}$}
}
\startdata
S-Value  & 0.02  & 0.26  & 0.021  & 0.526  & 0.195  & 0.186  & -0.005  & 0.072  \\ 
\ha\ Emission  & -0.317  & 0.564  & -0.051  & 0.209  & 0.005  & 0.003  & 0.0  & 0.0  \\ 
\ha\ Equiv. Wid  & 0.027  & 0.031  & -0.001  & 0.001  & -0.001  & 0.072  & -0.005  & 0.049  \\ 
CCF BIS  & 0.09  & 1.118  & -0.011  & 0.048  & -0.006  & 0.005  & -0.034  & 0.058  \\ 
CCF FWHM  & -0.005  & 0.534  & 0.001  & 0.022  & 0.002  & 0.001  & -0.008  & 0.118  \\ 
V\textsubscript{span}  & -0.076  & 0.567  & -0.007  & 0.009  & 0.016  & 0.018  & -0.007  & 0.006  \\ 
Bi-Gaussian Fit  & 0.082  & 1.498  & -0.005  & 0.015  & -0.006  & 0.038  & -0.008  & 0.154  \\ 
Skew Normal Fit  & -0.483  & 0.206  & -0.025  & 0.014  & 0.001  & 0.006  & 0.0  & 0.0  \\ 
\fdpca  & -0.001  & 2.418  & 0.0  & 0.775  & 0.0  & -0.017  & 0.0  & 0.255  \\ 
\gprn  &    &    &    &    & 0.0  & 0.054  & 0.0  & 0.362  \\ 
\scalpels  & -0.42  & 2.079  & -0.473  & 0.859  & 0.122  & 0.458  & 0.245  & 0.547  \\ 
\scalpels+\glom  & -0.206  & 2.31  & -0.202  & 1.217  & 0.143  & 0.496  & 0.271  & 0.571  \\ 
\ccfprime  & 0.182  & 1.76  & 0.0  & 0.222  & -0.01  & 0.032  & 0.124  & 0.18  \\ 
\fiesta+\glom  & 0.234  & 2.355  & 0.1  & 0.713  & 0.13  & 0.24  & 0.129  & 0.088  \\ 
CCF Linear Regression  & 0.001  & 2.607  & -0.176  & 0.56  & 0.196  & 0.297  & 0.065  & 0.196  \\ 
CCF LR + \ha  & -0.13  & 2.863  & -0.193  & 0.679  & 0.168  & 0.352  & 0.065  & 0.196  \\ 
CCF LR + Keplerian  &    &    & -0.015  & 0.484  &    &    &    &    \\ 
CCF Mask-\vald  & 0.202  & 1.336  & -0.01  & -0.001  & -0.142  & 0.021  & -0.002  & -0.029  \\ 
CCF Mask-BIS  & -0.231  & 1.421  & -0.289  & 0.505  & -0.292  & -0.134  &    &    \\ 
CCF Mask-RV  & -0.232  & 2.292  & -0.605  & 0.905  & -0.285  & -0.136  &    &    \\ 
\lblspec  & -0.182  & 2.36  & -0.226  & 0.85  & -0.027  & -0.098  & 0.088  & 0.261  \\ 
\lblrvel  & -0.421  & 2.46  & -0.354  & 0.51  & 0.122  & 0.286  & 0.101  & 0.33  \\ 
\lblboth  & -0.238  & 3.159  & -0.032  & 1.549  & 0.066  & 0.205  & 0.115  & 0.394  \\ 
\pwgp  & -0.022  & 2.132  & 0.179  & 0.857  & 0.232  & 0.374  & 0.251  & 0.376  \\ 
\dcpca  & 0.012  & 1.942  & 0.109  & 0.998  & 0.146  & 0.235  & 0.144  & 0.145  \\ 
\dcpca+\glom  & 0.027  & 2.368  & 0.108  & 1.125  & 0.147  & 0.241  & 0.144  & 0.111  \\ 
\horatio\ Self  & 0.123  & 2.86  & 0.117  & 2.01  & 0.629  & 1.041  & 0.496  & 0.644  \\ 
\horatio  & -0.165  & 0.374  & -0.172  & 0.251  & -0.1  & 0.076  & -0.041  & 0.053  \\ 
\hamlet  & -0.104  & 1.957  & -0.263  & 1.785  & -0.21  & 0.123  & -0.243  & 0.271  \\ 

\enddata
\tablecomments{All RMS values given in units of \ms.}
\end{deluxetable*}
\end{rotatetable*}

\noindent
reported results for with no data held out, as was done in \horatio\ Self.

Similarly, the use of \glom\ to co-model RVs and indicators almost always results in a lower RV RMS than the alternative (i.e., \scalpels+\glom\ returns a lower RMS as compared to \scalpels\ results and similarly with \dcpca+\glom\ as compared to \dcpca\ results).  In some cases, the use of \glom\ across methods returns RVs with a similar RMS (see \scalpels+\glom, \fiesta+\glom, and \dcpca+\glom\ for HD~101501 and likewise \fiesta+\glom\ and \dcpca+\glom\ for HD~10700).  This suggests that \glom\ is modeling out a similar degree of variation in a time series regardless of the indicator(s) it is given.

Methods that operate along very similar principles often return very different RV RMS results.  For instance, the different line-by-line methods (shown as light blue bars in Figure \ref{fig:rmsbar}), return RV RMS values that range from 77 to 102 \% the RMS of the originally provided RVs for HD~34411 and 35 to 73 \% the original RMS for HD~101501.  For most groupings, the HD~34411 results have the lowest spread while the HD~101501 results have the highest.  On the other hand, all methods that use \glom\ (i.e., \scalpels+\glom, \fiesta+\glom, and \dcpca+\glom) return similar resultant RMS values.  For HD~101501, all three methods return an RV RMS approximately 52 \% of the original RV RMS; the results of the other three targets have a percentage range of 10 to 20 \% between the lowest and highest RMS for each target.

We see here that the different methods do have a notable impact on the resultant RMS of the clean RVs, yet it is impossible to say from this one-dimensional metric what exactly is being modeled out by each method.  Just because a method is returning a lower RV RMS does not necessarily mean it is doing better at mitigating stellar signals specifically or is successfully preserving planet signals; this cannot be established from the RMS alone.

\begin{figure*}[tb]
\centering
\includegraphics[width=.96\textwidth]{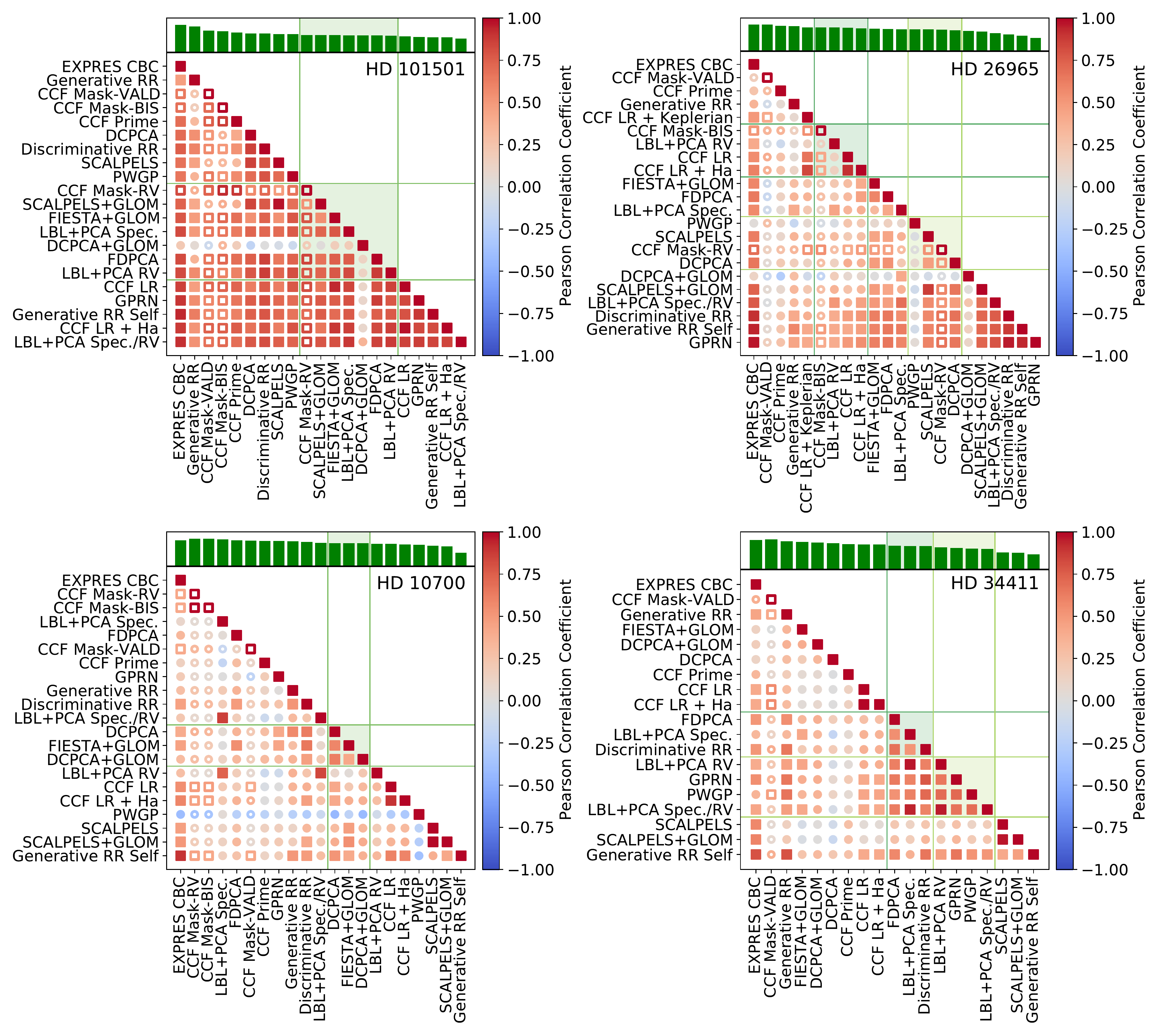}
\caption{Pairwise comparisons between the activity RVs of all submitted methods.  Each marker represents a pairing; the color of the marker gives the PCC between the activity RVs of two methods.  Methods that did not submit activity RVs (for which the difference between the original and clean RVs was used instead) are shown as marker outlines.  Significant PCC values (i.e., PCC $> 0.4$) are shown as squares while a PCC $<0.4$ is depicted with a circle.  The first column of each plot gives the PCCs of activity RVs with the provided CBC RVs.  The following rows/columns have methods ordered from top to bottom and left to right by decreasing total RMS, the same as how methods are ordered in Figure \ref{fig:rmsbar}.  At the top of each subplot in green is a scaled bar-graph of the resultant RMS for each method.  As recreated from Figure \ref{fig:rmsbar}, methods that returned similar final RMS values are highlighted in shades of green along with their associated correlation coefficients.}
\label{fig:methodCorr}
\end{figure*}

\begin{figure*}[hbt]
\centering
\includegraphics[width=\textwidth]{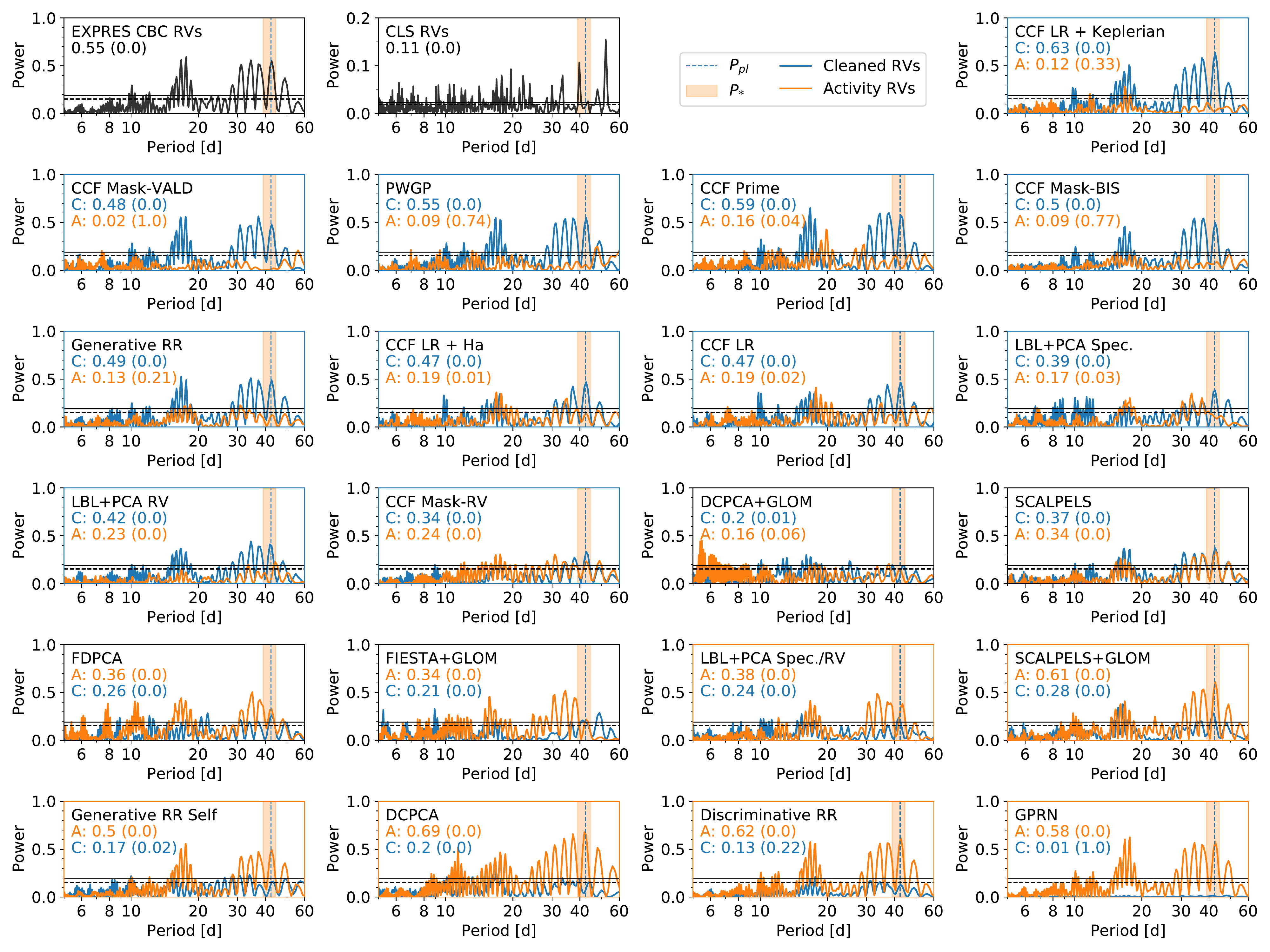}
\caption{Periodograms of RVs for HD~26965.  The leftmost two plots in the top row show the periodogram for the provided \expres\ CBC RVs (left) and over 30 years of RVs from the California Legacy Survey (CLS) on the right \citep{rosenthal2021}.  Periodograms of the clean (blue) and activity (orange) RVs are given for all twenty-one of the methods that submitted \essp\ results for HD~26965.  Horizontal dashed and solid black lines denote p-values of 0.1 and 0.01 respectively.  The proposed period for HD~26965~b, 42.38 days, is marked by a vertical, dashed blue line while the range given for the stellar rotation rate, 39-44.5 days, is shaded orange \citep{ma2018}.  The maximum power in this shaded region for both the clean (C) and activity (A) RV periodogram is given in the top-left corner of each subplot along with the corresponding p-value of the peak.  Methods are ordered left-to-right and top-to-bottom by the difference between the clean and activity periodogram peaks.  Subplots for methods resulting in a stronger peak with their clean RVs have blue axes; orange axes indicate a stronger periodicity in the activity RVs.  Methods with comparable peaks in their clean and activity RVs are shown with black axes.}
\label{fig:pva}
\end{figure*}

\subsection{Comparing Methods}\label{sec:results_corr}
Through the \essp, all teams were given the same set of \expres\ data to use with their respective methods and to model out stellar signals.  When working with real data, we do not know what stellar signals are present for each target.  Because the data are consistent among all methods, we would expect the stellar signal being removed to  be consistent between methods successfully modeling photospheric velocities.  Hence, the activity RVs for each method should be correlated with one another.

We perform a pairwise comparison of the activity RVs returned by each method.  For methods that do not naturally derive the RVs due to stellar signals, we approximate these activity RVs as the RV shift removed---i.e., we take the difference between the provided, CBC RVs and the submitted clean RVs to be the activity RVs.  For each pair of activity RV time series, which we expect to have a direct one-to-one correspondence, we use the Pearson correlation coefficient (PCC) to gauge the strength of correlation between the activity RVs derived by two different methods.

Figure \ref{fig:methodCorr} shows markers for each pair of methods colored by the PCC between the activity RVs each pair of methods returned\footnote{Note, we also generated an equivalent plot comparing the cleaned RVs with one another.  As all cleaned RVs are derived from the same provided RVs, nearly all clean RVs are significantly correlated with one another.  We felt this comparison provided less information than the comparison among the activity RVs and so chose not to include this figure.}.  The first column of each plot shows the PCC of each set of activity RVs against the provided CBC RVs.  A PCC of $>0.4$ with an associated p-value of $<0.05$ (square markers in Figure \ref{fig:methodCorr}) is considered statistically significant. This significance level was established with respect to the spread of PCCs returned from comparing series of randomly generated numbers of the same length as the RV data sets.

Comparisons between two methods that both submitted activity RVs are shown as filled in markers.  Comparisons involving activity RVs that were recreated as the difference between the provided RVs and submitted clean RVs are not filled in.  Only submitted methods are included; the results from classic linear decorrelation with standard activity indicators are not shown.

The top of each plot recreates a scaled bar graph of the final RMS of the clean RVs for each method.  These insets are meant to help associate each method with their final returned clean RV RMS.  Methods that returned similar final RMS values, as well as the relevant correlation markers, are highlighted in shades of green that mirror the shading in Figure \ref{fig:rmsbar}.

As expected, most PCCs are positive, but there is limited strong ($>0.4$) correlation.  Even methods returning similar RMS values to one another (i.e., markers close to the diagonal) are often not returning activity RVs that are significantly correlated with one another.  The methods returning the most similar RVs (as highlighted via green shading) are correlated for HD~10700 and HD~34411, but not for HD~26965.  HD~101501, the most chromospherically active of the four stars, has the most correlation amongst the activity RVs returned.

Methods returning lower clean RV RMS (i.e., methods closer to the bottom or further to the right of each subplot) are more likely to have activity RVs significantly correlated with other methods'.  These methods are even more likely to be significantly correlated with the provided \expres\ CBC RVs (see the first column of each plot).  If the derived activity RVs of these lower-RMS methods are subsuming much of the signal in the provided \expres\ RVs, then we would expect to see them show greater correlation with the provided RVs and all other methods that use the provided RVs as a starting point.

Variations on a method are nearly always significantly correlated with one another.  For example, the activity RVs returned by \scalpels\ and \scalpels+\glom\ are significantly correlated for all four targets.  The same is true for the three CCF Linear Regression variations (i.e., CCF LR, CCF LR + \ha, and CCF LR + Keplerian) and the three residual-regression based methods (i.e., \horatio, \horatio\ Self, and \hamlet).

Variations on line-by-line methods also agree with each other.  The results of CCF Mask-BIS and CCF Mask-RV are always correlated as are the results of \lblspec, \lblrvel, and \lblboth.  However, the activity RVs returned by these similar methods do not correlate strongly with each other.  Correlation with the \pwgp\ activity RVs is particularly lacking in the case of HD~26965 and HD~10700, where they are not correlated with the activity RVs returned from any other method.

The results of the \dcpca+\glom\ method are only correlated with the results of the \dcpca\ method for HD~10700 despite both methods being informed by the same indicator.  The \dcpca+\glom\ activity RVs are not correlated with the activity RVs of any other method for HD~101501 and HD~34411.  They are at most correlated with three other methods for the other two targets.

\begin{deluxetable*}{l | l l l }[bt]
\tabletypesize{\scriptsize}
\tablecaption{Method Philosophies \label{tab:philosophy}}
\tablehead{ Method & 
\multicolumn{1}{c}{Metric} &
\multicolumn{1}{c}{Mitigation} &
\multicolumn{1}{c}{Separation}
}
\startdata
\glom &  & Multi-Dimensional GP Modeling &  \\
\fdpca &  & Commonalities in Fourier Space &  \\
\gprn &  & GP Neural Net Modeling &  \\
\hline
\scalpels & PCA Amplitudes (CCF) &  & Shape/Shift-Driven RVs\\
\ccfprime & GP Model Coefficients &  & Shape/Shift-Driven RVs\\
\fiesta+\glom & Fourier Model Coefficients &  &  \\
CCF Linear Regression &  &  & Shape/Shift-Driven RVs\\
\hline
CCF Masks &  &  & Variable/Stable Lines\\
\lblspec &  &  & Variable/Stable Lines\\
\lblrvel & PCA Amplitudes (\lbl\ RVs) &  &  \\
\pwgp &  &  & Variable/Stable Lines\\
\hline
\dcpca & PCA Amplitudes (Spectra) &  &  \\
\horatio &  & Regression w/ Spectral Residuals &  \\
\hamlet &  & Regression w/ Spectral Residuals &  \\
\enddata
\end{deluxetable*}

\subsection{HD~26965 Results}\label{sec:res26965}

One of the hopes of including the HD~26965 data as an \essp\ target was to gain a deeper understanding of the $\sim$40 day, periodic signal.  This period had previously been associated with both the stellar rotation rate of the star and with a potential orbiting planet with a RV semi-amplitude of 1.8 \ms\ \citep{diaz2018, ma2018, rosenthal2021}.

For each of the submitted methods, we compare the periodogram of the clean RVs and the activity RVs, as shown in Figure \ref{fig:pva} in blue and orange respectively.  We also include periodograms of the provided \expres\ RVs and all RVs from the California Legacy Survey (CLS) \citep{rosenthal2021} in the top row for reference.  We focus on the power associated with periodicities between 39 and 44.5 days, the proposed stellar rotation rates for HD~26965, which bookend the proposed 42.38 day planet period \citep{ma2018}.  The maximum power in this period range along with the corresponding p-value is given in the top-left corner of each subplot.  Note that the \expres\ data peaks at 42.67 days, close to the proposed planet period.  The CLS data, on the other hand, peaks at 41.52 days, slightly lower than the proposed planet period, and features a much higher peak at 52.13 days.

Methods with more power (within the highlighted period range) in the clean RV periodogram are shown in blue subplots while methods with more power in the activity RV periodogram is shown in orange.  Four methods either have no significant peaks for those periods or return similar power in both the clean and activity periodograms (black axes).

Six out of twenty methods subsume the $\sim$40 day period in their stellar signal model while eleven of the methods produced clean RVs that still contain the $\sim$40 day period.  Five of the six methods that attribute the signal to stellar variations returned the five lowest RMS values for their clean RVs.  In these cases, almost all the variation in the provided RVs was modeled out as being due to stellar signals.  As we saw with the lack of correlation between activity RVs from different methods in Figure \ref{fig:methodCorr}, here we see again that the different methods do not agree on what signal is due to stellar variation and what can be attributed to an orbiting planet even for a prospective signal that is as large as 1.8 \ms.

\section{Summary}\label{sec:summary}
By using \expres\ data as a test bed for several different methods, the \essp\ is able to make a direct comparison between the results of twenty-two methods (including method variants) for disentangling stellar signals from true center-of-mass shifts.  Methods returned clean RVs, with stellar signals removed, and where appropriate activity RVs, which capture the variation that was removed.

The different methods varied in the type of data read in, metric for the presence of photospheric velocities, and mitigation of these signals once detected.  We compared method results based on the total and nightly RMS of the returned clean RVs, agreement between returned activity RVs, and conclusions with regards to the HD~26965 prospective planet.

\subsection{Categories of Methods}
Submitted methods for disentangling stellar signals operate along three broad lines.  Some methods innovate on the idea of activity indicators and use different models to derive a metric for the amplitude of the stellar signal present in an observation.  Other methods instead use such indicators and construct models for mapping this stellar signal amplitude measurement to the appropriate RV correction.

The last category of method separates the data into components that inform the true bulk shift of the star and components that add variability.  For instance, line-by-line methods separate variable lines from more stable lines that are assumed to be a better tracer of the true bulk shift of a star.  Many of the methods that model the CCF determine the shape-driven component of the measured RVs as opposed to the shift-driven component.

Table \ref{tab:philosophy} summarizes all submitted methods along these three lines.  Variations on the same method idea are not included.  Some methods naturally produce a metric as well and so operate along more than one of the three lines.

\subsection{Method Results}
The historical standard where RVs were linearly decorrelatied against activity indicators rarely changes the resultant RV RMS significantly.  This method of mitigating stellar signals is not sufficient in an EPRV context.

Most of the submitted methods reduce the RV RMS for all targets.  However, no method is able to completely model out the contribution from stellar signals.  \expres\ data of quiet stars exhibit an RMS of 0.5-0.8 \ms; no method was able to reduce the RV RMS to less than 1.2 \ms\ except for the \gprn\ method for HD~26965 only\footnote{We do not consider the results of the \horatio\ Self method here as it is not considered to be statistically rigorous.  This result was mainly included as a test of the importance of incorporating cross validation into model construction.}.

The reduction in RV RMS for method results relative to other methods changed from target to target.  HD~101501 and HD~26965 saw the most variation in relative method performance; a few methods returned much lower relative RMS values for HD~101501 and HD~26965 than they did for other targets.  HD~101501 is the most chromospherically active of the four targets.  HD~26965 was complicated by a proposed planet signal very close to the measured rotation rate of the star.  Relative method RMS was much more consistent between HD~10700 and HD~34411.  The change in behavior between the different stars hintst that methods may perform differently depending on the amplitude of the stellar signal or dominant type of variation exhibited by different stars.  This may also contribute to the lack of correlation seen between method results for the same star.

The average intra-night scatter changes very little, but does increase for some methods.  Whether the INS increases or decreases can also change for different targets with the same method.  We do not expect the magnetic field of a star to change on the timescale of a single night and even less so for consecutive observations taken on the same night.  This means that any signal from magnetically-driven stellar variability should be nearly the same for all observations taken within a night.  Methods that increase the INS may benefit from incorporating this constraint.

The activity RVs returned by the different methods often do not agree with one another.  All methods were used on the same data set and so should be capturing the same stellar signal.  Of course, different methods may have varying levels of success in modeling the observed stellar signal or be more/less sensitive to different types of photospheric velocities.  For methods that did not provide activity RVs, we instead compared constructed activity RVs from taking the difference of the originally provided CBC RVs and the submitted cleaned RVs.  These constructed activity RVs may be less correlated with methods that directly generated activity RVs as there is no guarantee such a construction will contain only stellar signals.  If the method modeled out more variation than just due to stellar signals, those variations will persist in the constructed activity RVs.

Some of the methods, most notably \dcpca+\glom\ for all targets and \pwgp\ for HD~26965 and HD~10700, are not correlated with the activity RVs of any of the other methods.  In the case of \dcpca+\glom, it is interesting to note that the \dcpca\ method results do not have the same issue, suggesting the \glom\ implementation for this method resulted in the lack of correlation.  It would be interesting to investigate why the \pwgp\ results exhibit no correlation for only two of the four stars.  This split behavior could be due to a systematic difference between the activity RVs being compared that are emphasized for some targets over others.  For example, methods may differ in what exactly was returned for the activity RVs, whether instrumental or telluric variation was also fit for, and many other implementation specifics.

The lack of agreement between methods makes it difficult to confidently state what signal is being modeled and removed by each method to result in the observed RMS reduction.  It is also a demonstration of why RV RMS alone is an incomplete metric for method performance, as is fails to establish the nature of the signal that a method is capturing.  Many methods invoke tunable parameters to control how much variability is fitted out as due to stellar signals.  In optimizing these parameters, the resultant RV RMS should be used as a goodness metric with caution and other metrics should be considered (see \S \ref{sec:future_methods} for some discussion).

The disagreement among methodsis further highlighted with the HD~26965 results.  Whether or not HD~26965 hosts a planet changes depending on the method used.  Some methods model out the $\sim$40 day period as due to photospheric velocities while others attribute that periodicity to true center-of-mass shifts.  Many methods found it difficult to account for the planet signal as well as the close-by stellar rotation rate.  A test with an injected Keplerian signal of known period would give clearer results.

Results for HD~101501 were most correlated with other method results.  This suggests that the inferred corrections are more similar for stars with a larger amplitude of magnetic activity.  The difference in performance could also be due to other stellar parameter specifics.  With more test cases, it may become clear whether methods tend to perform better depending on the spectral type of the star, expected activity, or other stellar properties.

\section{Discussion}\label{sec:discussion}
An increasing number of EPRV instruments are coming online and returning sub-meter-per-second single-measurement precision \citep[e.g.,][]{pepe2013, jurgenson2016, schwab2016, carmona2018, seifahrt2018, gilbert2018, blackman2020, petersburg2020, mascareno2020, pepe2021} with many more optical and infrared spectrographs being commissioned, built, or planned \citep[e.g.,][]{szentgyorgyi2014, thompson2016, bouchy2017, gibson2018}.  The impressive engineering feat of these different instruments is opening up a new regime of extremely stable and precise spectroscopic data.  However, each of these instruments and the data they take will have to contend with added RV scatter due to chromospheric velocities unless we can mitigate these effects to below 50 \cms\ levels.  None of the methods presented in this paper were able to consistently achieve that across the data sets provided.

Though there is no one single method clearly performing the best, this collection of methods and results brings clarity to the approaches and assumptions that define the current state of the field.  Here, we will highlight some of the commonalities between methods.  From this, we derive suggested future directions both for method development and continued coordinated data releases like the \essp.

\subsection{Common Approaches and Assumptions Between Methods}
The choice of input data changes the information made available to each method.  For instance, indicator-driven methods will only be able to pick up on stellar signals that are tracked by the indicators used.  Similarly, CCF-based methods will only be able to account for variations that are present in the CCF.  None of the methods made use of the provided photometry.  Some methods were not able to use the photometry because it was not simultaneous with the RVs.  Both CCF-based methods and methods that use the full spectra can only account for line-shape variations at the level of the resolution of the spectrograph.  Higher resolution data will contain more information about line-shape changes.

Currently, methods are tracing stellar signals using either global activity indicators, spectral line-shape variations, or increased scatter.  It is worth considering and perhaps attempting to simulate whether stellar signals may manifest in a way that is not captured by current metrics and therefore are not being modeled by existing methods.  We know that indicators are imperfect.  Stellar signals may manifest as a shift in addition to line-shape changes. Taking increased scatter to be synonymous with stellar variation/activity is a dangerous parallel as we have seen that a reduction in RMS does not necessarily equate with mitigating a stellar RV component.

Methods that measure only shape-driven changes will miss stellar signals that manifest as a shift.  Indeed, no method will be able to mitigate shifts caused by stellar signals unless they can be disentangled from center-of-mass shifts.  As just one example, stellar oscillations may cause all spectral lines to shift like they would in the case of bulk motion of the star.  This shift could maybe be disentangled from center-of-mass shifts by a method that links the shifts due to oscillations with the expected associated changes in stellar parameters (e.g., temperature, luminosity, log $g$, etc.), which would exhibit different or no changes due to an orbiting planet.  Future work needs to be done to investigate how stellar signals may manifest as shifts and what metrics exist to distinguish such a shift from bulk, center-of-mass shifts due to planets.

Many methods assume a diversity of activity states or, more specifically, that the effects of stellar signals captured in a data set span a large range of amplitudes.  Methods that model the effects of stellar signals using PCA assume that stellar signals are the primary source of variation and are therefore traced by the first few/several principal components.  Using correlations with indicators or increased scatter to determine the presence of stellar signals is also helped by having a large range of activity states sampled.

Template CCFs and spectra are used as a point of reference for many methods.  Methods varied in whether they used the mean, median, or optimization (e.g., \wobble) to construct this template.  These templates are used to highlight variations away from the template, which are then attributed to the presence of stellar signals.  The ideal template will not carry any significant variations due to stellar signals so that it can be used as a reference to isolate those variations in each individual observation.  Constructing a mean or median template CCF/spectrum and using it to highlight changes due to chromospheric velocities therefore assumes an even sampling of activity states that will average out.  It would be worthwhile to investigate how dependent method results are on the template used.  Methods could be run using different subsets of the data to construct the needed template and see how much the results vary.

On a similar note, methods that ascribe deviations from a Gaussian fit as an indication of stellar signals inherently assume that the line shape and CCF shape is well-described by a Gaussian fit.  We see no evidence otherwise with the \expres\ data, for which great pains were taken to stabilize the instrument LSF across the detector.  Were this not the case, however, any instrumental deviations from a Gaussian profile could be mistaken for shape changes due to stellar signals.

Many methods make the assumption that Gaussian processes and principal component analysis are good models for stellar signals.  Different methods, however, implement GPs and/or PCA in distinct ways.  For instance, \glom\ uses a GP to model a time series while the \gprn\ model uses GPs to define a neural net framework.  \ccfprime\ forms a basis out of the derivatives of a GP model.  In each case, a GP is implemented towards different ends and  requires different assumptions of the appropriate kernel, hyper parameter priors, etc.

PCA can be used to construct a variation specific basis or as a measure of the amplitude of variation.  Roughly speaking, the distinction can be made based on what aspect of the PCA is used.  Some methods (e.g., \fiesta, \dcpca, \lblrvel, \scalpels+\glom) use just the amplitudes for each component derived from PCA as a measure of variation and therefore of photospheric velocities.  Other methods (e.g., \fdpca, \scalpels, \lblspec) make use of the principal components themselves to denoise spectra or model the RV shift tied to the variation being modeled by the PCA.  As PCA is agnostic to the source of the variation, and cares only about the amplitude, implementations of PCA may also be picking up on variations from the instrument, the extraction, tellurics, etc.\ which are not stellar in nature (though important to correct for nonetheless).  This may also be a cause of the lack of agreement we see in the activity RVs returned by different methods.

Derived RVs are often used to align CCFs/spectra (for example for template construction), thereby implicitly assuming that true center-of-mass shifts from orbiting planets have been or can be removed leaving only stellar signals.  We know, however, that these measured RVs are swayed by stellar signals.  Methods should consider iterating with clean RVs produced by methods given different results, provided we are confident the corrections are truly removing only stellar signals \citep[e.g.,][]{cretignier2021}.

Methods mostly operate under the self test framework, meaning all data is used to construct the model with no built-in cross-validation framework, unless otherwise stated.  From comparing the results between \horatio\ and \horatio\ Self, we saw that the \horatio\ Self method always returned a lower RMS but the returned activity RVs were not even linearly correlated with the activity indicator used to guide the model.  This suggests that the \horatio\ Self model was over-fitted and absorbed signals that are not informed by the indicator, something the cross-validation aspect of \horatio\ guarded against.  Implementing leave-one-out, such as is done here by \hamlet and described for \scalpels\ in \cite{collier2021}, or other cross-validation tests, such as the framework for \horatio, should be a default of methods disentangling stellar signals when applicable in order to ensure the stability of the model being used.  Cross-validation tests are more effectively run on larger data sets.

\subsection{Future Directions for Methods}\label{sec:future_methods}
The reduction in RMS with the cleaned RVs of the different methods is encouraging, but with a one-dimensional metric of method performance, it is not clear what exactly is resulting in this reduced scatter.  This is especially worrisome given the lack of agreement between method results.  To progress, methods should be held to a higher level of interpretability.  Understanding what exactly methods are tracing will be helpful in developing them further and build confidence that potential planetary signals are preserved.

The new types of activity indicators being generated should be tried with the different methods that take indicators as input (i.e., as outlined in columns one and two of Table \ref{tab:philosophy} respectively).  For example, \glom\ is used with different generated indicators from \scalpels, \fiesta, and \dcpca\ here.  Rather than trying to find one, ``best" method as they are currently named, we should instead be testing all combinations of metrics and mitigation strategies.  This will more fully explore the parameter space and help establish whether it is a metric or mitigation method that is the main driver of a method's performance.  Ultimately, this will allow for a better informed down selection of methods and frameworks worth further investigation.

Methods modeling shape changes may benefit from implementing low-pass filtering tuned to the resolution of the spectrograph.  The information content in a spectra is limited by the resolution of the spectrograph.  Filtering out effects above this level would prevent methods from being swayed by higher-frequency variations than is allowable by the spectrograph resolution, which therefore must be due to noise.

Results of the different separation methods (i.e., methods outlined in column three of Table \ref{tab:philosophy}) should be compared with one another to see if any ground truth can be established.  For instance, all line-by-line methods work to identify lines that are more or less variable.  It would be informative to understand which lines the methods agree on and for which lines they differ.  Using physical information about the different lines, e.g., the line's element, transition specifics, formation level in the stellar photopshere, etc., can lend interpretability to these line-by-line methods and other methods that identify variation in the spectra.  It may also be useful to consider what commonalities are shared between methods that use the same input data (e.g., CCF, spectra, etc.) and whether there is a benefit to using one type of input over others.

Line-by-line methods have thus far primarily used scatter in returned RV, correlation with different activity indicators (classic or otherwise), and error of resultant RV to vet lines or chunks.  More advanced methods for vetting may be interesting to explore.  For instance, a periodogram of the RVs returned by individual lines or chunks could be used to vet for ones that show power at troubling periods, e.g., the stellar rotation rate, p-mode oscillation timescale, etc.  Clustering analysis may also be useful in identifying lines or chunks with similar properties and help link problematic retions with one another.

The axes of variation revealed by the different PCA methods could be picking up on the same variations.  Commonalities between methods lends significance to the variations captured, which could be traced back to effects we would expect from an understanding of stellar physics.  Different methods decomposing the CCF should have some commonalities even if the basis used varies greatly.

None of the methods analyzed here made use of the provided photometry, though such efforts exist and have shown success \citep[e.g.,][]{aigrain2012, cabot2021, roettenbacher2021}.  As an independent probe of activity on the stellar surface, photometry has proved useful for linking the signal being modeled with changes on a star's surface \citep{kosiarek2020}.  Incorporating photometric information into more methods would help with method interpretability by tying the modeled RV signals to a separate measure of activity.  Simultaneous photometry, which was not provided, is most immediately useful for current methods.  Teams also expressed a preference for space-based photometry, but a full investigation of the requisite photometric data quality is beyond the scope of this paper.

Currently, we do not have a good understanding of the precision or cadence of photometry needed to inform EPRV work.  Future research should work to understand the quantity/quality of the photometry needed to guide methods for disentangling stellar signals.  Current implementations of methods suggest that simultaneous photometry should be prioritized.

\subsection{Future Directions for Data Challenges}
Comparing methods with consistent, realistic data sets will grow increasingly important as \expres\ and other next-generation spectrographs continue collecting high-fidelity data.  Such comparisons will be more informative with better metrics for method performance.

For this report, we carried out only a few fairly simplistic tests using the relative RMS of the submitted RVs and correlation between the submitted RVs of different methods.  We have seen that RMS is not sufficient to capture exactly what a method is modeling out.  We tried comparing the returned activity RVs to different activity indicators (both classic and those derived by the submitted methods) using Spearman's rank correlation coefficient (SCC), but it was unclear what a lack of correlation meant.  No significant correlation could indicate over-fitting in the activity RVs, a fault in the activity indicators, and/or the existence of a more complicated relation between the activity indicators and activity RVs that is not captured by SCC (for example the two may be out of phase).

The field would greatly benefit from the development of more representative comparison metrics.  Such metrics should focus on diagnosing the extent to which various methods are specifically capturing the effects of stellar signals.  Ground truth is not known with real data, so more advanced metrics should leverage the fact that all methods are probing the same underlying stellar signal, although to various levels of precision.  For instance, invoking a periodicity dependence or expectation for the effects of stellar signals beyond increasing scatter would be a good start.  Establishing a standard suite of assessments for all methods will help place old and new methods in context.

Interpretability is easier to establish when there is a known ground truth---i.e., what the stellar signal is expected to be, and what is a true center-of-mass shift.  One such test would be to inject simulated, center-of-mass shifts into real data at the spectral level from which all CCFs, RVs, and activity indicators are derived\footnote{See \cite{collier2021} for an example of injected shifts at the level of the CCFs.  See \cite{dumusque2016} for a discussion of injecting planet, stellar, and instrumental variations at the level of the RVs.}.  Methods that are truly only picking up on stellar signals will preserve these injected center-of-mass shifts.  The most informative simulations will be shifts of the magnitude similar to the RMS of the data and at periods near the stellar rotation rate or its harmonics, as these signals will be the hardest to disentangle.

A kind of ground truth is also known for well-characterized systems, the prime example of which is our Sun.  The Sun remains one of the few stars for which we can definitively remove all planet shifts\footnote{Here we are assuming the RV signal from the proposed Planet 9 would be below the white noise level; most constraints on Planet 9's orbit correspond to an RV semi-amplitude of $\sim$4 \cms\ with a period of $\sim$ 7,500 years \citep{batygin2016, batygin2019, millholland2017, brown2021}}.  Any remaining variation in the solar spectra will be from stellar signals or instrumental variation.  We are also able to trivially image the surface of the Sun and directly see changes.  With several solar telescopes expected to accompany next-generations instruments coming on line, simultaneous observations using different instruments along with photometry and surface maps will help isolate stellar signals from unique instrumental variation.  Dense sampling and high cadence will additionally be immensely more achievable for the Sun than with other stars.

At the same time, the field should be careful not to become overly reliant on solar data or simulations constructed with exclusively solar data.  Stellar signals and their spectral manifestations differ for different types of stars.  Additionally, stellar data is free from the $\pm 20$ \kms\ barycentric corrections that affect other stars, which will shift stellar lines across different telluric lines and across different detector locations.  It is necessary to build up the ability to convincingly simulate or thoroughly characterize stellar signals that arise from a range of spectral types to ensure that method performance is universal.

Future data can serve as the truest validation set for methods trained on the already provided data and be used to uniformly diagnose the generality of the models constructed by each method.  Carrying out this useful test will require data sets that can be separated into a large enough training set to inform all different types of methods and, correspondingly, a large enough validation set to confirm the model results.  More data will also likely sample a greater range of activity states, resulting in additional variation in the observed spectra that will improve the performance of all methods.

The existing data along with any future data can be used to empirically determine data requirement limits for methods.  We can synthetically degrade the data to establish how method performance depends on different aspects of the data quality.  For example, in addition to total number of data points, the cadence of the data (e.g., $n$ observations in a month vs. $n$ observations over a year) or nightly sampling (e.g., three observations per night or only one) can be adjusted.  The SNR or the resolution of the observations can also easily be degraded.

There are currently several data pipelines and methods for extracting spectra and removing instrumental signals \citep{petersburg2020, zhao2021, cretignier2021}.  It is worth considering the effect different extraction pipelines may have on the ability to remove stellar signals.  Method performance could change depending on the degree to which instrument variations are addressed, the wavelength calibration, whether the echelle orders are merged, the continuum normalization, etc.

Similarly, adjusting CCF masks and construction methods is an area of ongoing research, as we saw with the various CCF Mask methods.  The best CCF line list, mask window, and pipeline differs for different stars but may also change for different use cases.  For instance, the method results given here chose quiet lines to return quiet RVs, but there may be a use case for choosing the identified variable lines to construct a CCF mask meant to highlight the signatures of stellar variability.  Though we requested that all CCF methods use the provided CCFs for this report, exploration is warranted as to how different CCFs may change the results of these methods.

Currently, the focus of many methods and indicators lie in tracing activity features or magnetic field strength; less emphasis is placed on inherent stellar variability, such as p-mode oscillations or (super)granulation.  Pulsations and changes in granulation pattern persist on the timescale of minutes while supergranulation has a timescale of hours to days.  Pulsations may cause lines to shift rather than change in shape.  These types of variation will have a different diagnostic than activity features.

Before we can disentangle the effect from granulation, we must understand it.  This will require very densely sampled observations at high resolution.  Given the timescale of pulsations and (super)granulation, the ideal data set will have very dense sampling over the course of a night for four to five  consecutive nights in order to capture both short-term pulsations and granulation variations and potentially day-long supergranulation effects.

\section{Conclusions}\label{sec:conclusion}
Twenty-two different methods (including variations) were tried on \expres\ data to produce a consistent comparison of method results on data that are representative of extreme-precision instruments.  Since the ground truth is not known with real data, method performance must be established relatively.  The methods tested return lower RMS values than the classic linear decorrelation methods in nearly all cases.  Though \expres\ data of quiet stars regularly return RMS values of 0.5-0.8 \ms, no method is yet consistently reducing the RMS of more chromospherically active stars to sub-meter-per-second levels across all four stars (\S \ref{sec:results_rms}).  Lack of agreement between the signals being modeled out by different methods makes it difficult to determine exactly what variation is being modeled and whether it truly is stellar in origin (\S \ref{sec:results_corr}).

Current and future methods should consider:
\begin{itemize}
    \item increasing method interpretability in order to establish the source of the signals being picked out by the method, 
    \item ensuring models are appropriately general by implementing cross-validation tests, 
    \item iterating when aligning CCF/spectra with derived RVs, and
    \item making methods robust to the assumption that a large range or equal distribution of activity states is covered within the data set.
\end{itemize}
Methods currently work at identifying the presence of stellar signals by using either a derived activity indicator, changes in line shape, or increased scatter.  Future investigation is warranted as to whether those diagnostics are comprehensive and what manifestations of stellar signals are currently being missed.

None of the methods made use of the provided photometry, which is non-simultaneous and ground-based (\S \ref{sec:aptData}).  Previous work establishes photometry as a useful, independent measure of stellar surface changes for mitigate stellar signals, but the exact quantity/quality of the photometry needed remains an open question.  Space-based, simultaneous photometry would be easiest to incorporate into the current implementation of methods.

Next steps for establishing method performance include:
\begin{itemize}
    \item developing more holistic metrics for how well a method disentangles stellar signals, 
    \item cross-pollinating methods that generate activity indicators with methods that are informed by indicators, 
    \item comparing and contrasting results of similar methods, e.g., \lbl\ methods, derived PCA components, GP hyperparameters, etc., 
    \item testing methods on well-characterized systems, e.g., solar data, dynamically packed planetary systems, data with injected Keplerian or stellar signals, etc., and
    \item testing methods on data sets from \expres\ and other state-of-the-art RV instruments (e.g., \espresso, \neid, etc.) degraded in terms of SNR, resolution, observing cadence, etc.
\end{itemize}
Note that care must be taken when injecting a Keplerian signal to ensure telluric lines are not also shifted.  Injecting stellar signals will require developing simulations capable of faithfully reproducing all flavors of stellar variability and activity across different stellar types.

The design of RV surveys should consider whether to prioritize phase coverage of potential planets or to prioritize fully characterizing the effects of stellar signals.  An EPRV data set that fully resolves all timescales of stellar signals, including the shortest, minute-long timescales, is needed to completely understand the effects of chromospheric velocities on spectra.  Such a data set for a Sun-like star would likely need to span 4-5 consecutive nights with at least 2-3 hours of continuous, densely sampled observations per night.

While progress is being made in mitigating stellar signals, more work remains to be done.  We will not be able to successfully detect Earth-like planets until photospheric velocities from inherent stellar variability and activity features can be disentangled to below the 50 \cms\ level.

\vfill

\facilities{LDT, TSU:APTs}

\software{SciPy library \citep{scipy}, NumPy \citep{numpy, numpy2}, Astropy \citep{astropy2013,astropy2018}.}

\acknowledgements
We gratefully acknowledge the anonymous referee for their helpful and thorough comments.
These results made use of the Lowell Discovery Telescope at Lowell Observatory. Lowell is a private, non-profit institution dedicated to astrophysical research and public appreciation of astronomy and operates the LDT in partnership with Boston University, the University of Maryland, the University of Toledo, Northern Arizona University and Yale University.

LLZ gratefully acknowledges support from the NSF GRFP under Grant No. DGE1122492 and from the Green family.  DAF acknowledges support for the design and construction of EXPRES from NSF MRI-1429365, NSF ATI-1509436 and Yale University. DAF gratefully acknowledges support to carry out this research from NSF 2009528, NSF 1616086, NSF AST-2009528, the Heising-Simons Foundation, and an anonymous donor in the Yale alumni community.  This work was partially supported by NASA Exoplanet Research Program Grant \#80NSSC18K0443 (DAF, EBF, AW, JZ).  This work was supported by a grant from the Simons Foundation/SFARI (675601, E.B.F.).  This research was partially supported by Heising-Simons Foundation Grant \#2019-1177 (E.B.F).  The Center for Exoplanets and Habitable Worlds is supported by the Pennsylvania State University and the Eberly College of Science.  This work has made use of the VALD database, operated at Uppsala University, the Institute of Astronomy RAS in Moscow, and the University of Vienna.

ACC acknowledges support from the Science and Technology Facilities Council (STFC) consolidated grant number ST/R000824/1 and UKSA grant ST/R003203/1.
AM acknowledges support from the Cambridge Kavli Institute Fellowships.
H.M.C. and M.L. acknowledge support from the UKRI FLF grant MR/S035214/1.
Work by SDR, JH, and VRD was supported by Bartol Research Institute. VRD received additional support from the University of Delaware Summer Scholars Program.  The Sidera team gratefully acknowledges contributions from Catherine Lembo.
JDC, JPF and PTPV were supported by the following grants, awarded by FCT - Fundação para a Ciência e Tecnologia and FEDER through COMPETE2020: UIDB/04434/2020; UIDP/04434/2020; PTDC/FIS-AST/32113/2017 and POCI-01-0145-FEDER-032113; PTDC/FIS-AST/28953/2017 and POCI-01-0145-FEDER-028953.
JPF is further supported in the form of a work contract funded by national funds through FCT with reference DL57/2016/CP1364/CT0005.
SA, BK and OB acknowledge support from the European Research Council (ERC) under the European Union’s Horizon 2020 research and innovation programme (Grant agreement No. 865624). NZ is supported by studentship no. 1947725 under Grant Code ST/N504233/1 from the UK Science and Technology Facilities Council (STFC).
X.D. and M.C. acknowledge that this project received funding from the European Research Council (ERC) under the European Union's Horizon 2020 research and innovation programme (grant agreement SCORE No 851555).  X.D. and M.C. also recognise that this work has been carried out within the framework of the NCCR PlanetS supported by the Swiss National Science Foundation.
ZLD acknowledges the National Science Foundation Graduate Research Fellowship under Grant No. \#1745302. ZLD also acknowledges the generous support from the UT Office of Undergraduate Research Fellowship, the TIDES Advanced Research Fellowship, Dean’s Scholars, and the Junior Fellows Honors Program.

G. W. H. acknowledges long-term support from NASA, NSF, Tennessee State University, and the State of Tennessee through its Centers of Excellence program.
RMR acknowledges support from the Yale Center for Astronomy \& Astrophysics (YCAA) Prize Postdoctoral Fellowship and the Heising-Simons 51 Pegasi b Postdoctoral Fellowship.

\bibliography{main}

\appendix

\section{In-Depth Descriptions of Methods That Use RVs and Classic Activity Indicators as Input}

\subsection{\glom}\label{sec:glom}
\glom, developed by members of the PennState Team, is a software package for joint GP modeling of several parameters, such as Doppler shifts along with one or more activity indicator time series \citep{gilbertson2020-02}.  The model is based on the assumption that all time series can be modelled using a latent variable $G(t)$, which is described by a Gaussian process and a covariance function $\gamma$.  The \glom\ implementation can also incorporate a non-zero mean function, $m_n(t)$ for each set of variables being modeled.

RVs and activity indicators are modeled together using the latent GP $G(t)$, its derivatives, and this mean function.  For N total number of parameter time series, the framework is as follows:
\begin{equation}
\begin{aligned}
    q_0(t) &= m_0(t) + a_{0,0}G(t) + a_{0,1}\dot{G}(t) + a_{0,2}\ddot{G}(t) + \epsilon_{0}(t) \\
    q_1(t) &= m_1(t) + a_{1,0}G(t) + a_{1,1}\dot{G}(t) + a_{1,2}\ddot{G}(t) + \epsilon_{1}(t) \\
    & \hspace{1.75 mm} \vdots \\
    q_N(t) &= m_N(t) + a_{N,0}G(t) + a_{N,1}\dot{G}(t) + a_{N,2}\ddot{G}(t) + \epsilon_{N}(t)
\end{aligned}
\end{equation}
Each $q_{n}(t)$ is the time series of the variables being modeled.  The variables $a_{n,0}$, and $a_{n,1}$, where $n=1,...,N$, are free parameters and $\epsilon_n(t)$ represents measurement uncertainties.

GP models are a powerful tool for modeling stochastic behavior and therefore very apt for modeling photospheric velocities.  However, they are liable to vacuum up all signals in a data set including, for instance, planet signals.  By modeling several time series simultaneously, this method places constraints on the GP model by incorporating the information from activity indicators into the GP modeling.  This guides the model to only pick up on signals that can be tied to the provided indicators.  Introducing indicators into the modeling increases the size of the correlation matrix, making the method more computationally expensive.

The method requires RVs and corresponding indicator time series for each observation.  Photometry can be used to establish a constraint on the stellar rotation period of the target.  \glom\ is incorporated as a part of many submitted methods that generate different indicators of activity.

The success of the method is dependent on the sampling of the data, which should be relatively close in time, and the appropriateness of the chosen GP kernel.  It would be better to have less observations but a denser sampling throughout the characteristic timescale of the signal being modeled (i.e., the stellar rotation rate).  The GP model adopts a quasi-periodic kernel along with constant offset and jitter terms for each time-series.  Some care must be taken in choosing the priors for the GP hyper-parameters, which will change for different data sets.

\subsection{\fdpca}\label{sec:fdpca}
Fourier Domain Principal Component Analysis, submitted by the Sidera team, detects common patterns in the Fourier coefficients of RV and activity-indicator time series and uses this to predict the stellar signal component of the RV.  Moving to the Fourier domain allows the method to identify and remove correlated signals even if they are out of phase.  The power of this method comes from identifying coherences between the provided indicators and the RV measurements.

First, the non-uniform Fourier transforms of all activity-indicator time series and RVs are computed.  Next, the activity-indicator Fourier series are scaled so that they have unit variance in the time domain.  The Fourier series for each activity indicator are then stacked into a matrix to form a set of explanatory variables for the RV Fourier series:
\begin{equation}
\begin{bmatrix} \Re(\mathcal{F}\{\mathrm{H}\alpha {\rm EW} \}) & \Im(\mathcal{F}\{\mathrm{H}\alpha {\rm EW} \}) & \Re(\mathcal{F}\{{\rm CCF \; FWHM} \}) & \Im(\mathcal{F}\{{\rm CCF \; FWHM} \}) & \cdots \\
\end{bmatrix}
\end{equation}
where $\Re(\mathcal{F})$ and $\Im(\mathcal{F})$ are the real and imaginary parts of the Fourier transform, respectively.  The matrix is then run through PCA. \footnote{\fdpca\ was implemented with the following \texttt{python} packages: Flatiron Institute's \texttt{finufft} for non-uniform FFTs, \texttt{sklearn.preprocessing.StandardScaler} for scaling Fourier series to have unit variance, \texttt{sklearn.decomposition.PCA} for the PCA, and \texttt{sklearn.linear\_model.LinearRegression} for the linear regression.} 

With activity principal components in hand, the real and imaginary parts of the RV Fourier series can be regressed onto these principal components.  The regression coefficients are used to determine the proportion of the RV Fourier series that is related to the activity indicators.  This measures the chromospheric contribution to the RV Fourier series and can then be inverse transformed back into the time domain to find the stellar signal correction needed for each RV.  Parseval's theorem is used to recover the correct variance of the RV activity contribution.

Implementing this method requires RVs and indicators taken at the same time stamps.  In order to use this method to measure a signal, the observations must completely cover the phase of the signal.  For example, to capture the effects of a rotating activity feature, the observations must completely sample the star's rotation.  It is not just a question of dense sampling of observations, the observations must cover the entire phase range.

As with all methods that invoke PCA, there is always the question of how many principal components to incorporate.  For the results presented here, principal components were included until 95\% of the total variance was captured.

\subsection{\gprn}\label{sec:gprn}
The Gaussian Process Regression Network method, submitted by the Porto team, adaptively combines GP models to jointly describe variations in the RVs and activity indicators.  The structure of a \gprn\ share some similarities to an artificial neural network, with independent GPs acting as both nodes and weights.  Following the work of \cite{Wilson2012_GPRN}, a GPRN can model a function \textbf{y}(x) as 
\begin{equation}
\label{eq:GPRN}
\textbf{y} (\textbf{x}) = \textbf{W} (\textbf{x}) \textbf{f} (\textbf{x}) + \sigma_y \textbf{z}(x).
\end{equation}

On this network \textbf{f}(x) and \textbf{W}(x) are independent GPs,
\begin{equation}
\label{eq:NodesWeights}
\begin{aligned}
    & f_j(x) \sim \mathcal{GP}(0, k_f)~\text{for}~j = 1, ..., q, \\
    & W_{ij}(x) \sim \mathcal{GP}(0, k_w)~\text{for}~i = 1, ..., p~\text{and}~j = 1, ..., q.
\end{aligned}
\end{equation}

This framework is capable of accommodating noise correlations between multiple output variables as well as input dependent signals, length-scales, and amplitudes.  It leads to heavy tailed predictive distributions.

The method requires RVs and activity indicators as inputs, where each RV measurement must have a corresponding activity indicator taken at the same time stamp.  For instance, non-simultaneous photometry could not be used as an indicator.  The number of nodes and weights, as well as the associated co-variance functions, can be decided a priori or a posteriori based on marginal likelihood comparison.

In principle, each one of the GPs that form a node or weight of the regression network has its own set of associated hyper-parameters and respective priors.  However, it is possible to share hyper-parameters to reduce the number of free parameters, for example between the GPs acting as weights.  For the results presented here, only one node was defined by a GP with a quasi-periodic co-variance function.  GPs with squared-exponential kernels were used for the weights with no shared hyper-parameters.

\section{In-Depth Descriptions of Methods That Use the CCF as Input}
\subsection{\scalpels\ and \scalpels+\glom}\label{sec:scalpels}
\scalpels, submitted by the St.\ Andrews and PennState teams, makes use of autocorrelation functions to separate out Doppler shifts from shape changes that are attributed to stellar signals \citep{collier2021}.  The autocorrelation function of either the spectra itself or its CCF can be used.  In the velocity domain, the autocorrelation function is invariant to translation.  Projecting the measured velocity time series onto the principal components of the autocorrelation function isolates shape-driven shifts.  Because they are translationally invariant, these projected perturbations can be subtracted from the original velocities with the dynamical shifts preserved.

Applying the method requires either the spectra or the CCF to derive the autocorrelation function as well as the barycentric corrected time stamps, RVs, and RV errors for each observation.  From this, SCALPELS will output velocity variations that are driven by shape changes.  Subtracting out these shape-driven velocities leaves the true dynamical shifts preserved.

Since \scalpels\ operates in the wavelength-domain, it does not require any information about the star's behavior (i.e., rotation rate, pulsation timescale, etc.) nor does it need very dense sampling of the stellar rotation cycle.  Ideally, there should be at least 40 observations of a target over a full range of stellar activity states.  Observations taken at different activity states help the PCA of the autocorrelation function identify variations due to shape changes.

All \scalpels\ results presented here use the autocorrelation function of the provided \expres\ CCFs.  Results can vary with number of principal components incorporated.  The submissions given here used two principal components to minimize the risk of over-fitting.

The PCA results from \scalpels\ were also input into \glom, where the amplitudes of the principal components, i.e., the magnitude of the shape variation modeled in the CCF auto-correlation function, were used as activity indicators and modeled along with the RV shifts.  This process was run using the sum of two Mat\'{e}rn $\frac{5}{2}$ kernels for the latent GP model.

\subsection{\ccfprime}\label{sec:ccfprime}
The \ccfprime\ method, submitted by the \oxgen\ team, is an exploratory approach to decomposing the CCF by linearly modeling variations in each spectra's CCF using derivatives of a GP model.  A reference CCF is constructed by modeling the mean CCF of all observations using a GP with a square-exponential kernel.  Let this reference CCF be denoted by $C(v)$ where $v$ are the velocities at which the CCF is sampled. The quotient of each CCF against this reference CCF is then linearly modeled.

Let $c_i(v)$ denote the quotients of each CCF against the reference CCF, i.e., $c_i(v)=\frac{C_i(v)}{C(v)}$, where $i$ indexes over all exposures and $C_i(v)$ is the CCF for exposure $i$.  The linear model is then defined by the following equation
\begin{equation}\label{eq:ccfprime}
    c_i(v) = a_i + \Sigma_{k=0}^3 b_{ik}C^{(k)}(v)
\end{equation}
where $k$ corresponds to the different derivatives of $C(v)$ with respect to velocity.  In this case, $C^{(0)}(v) = C(v)$.  The parameters $a_i$ and $b_{ik}$ are the linear parameters of the model.

The first derivative term in equation \ref{eq:ccfprime} is sensitive to shift-induced variations on the CCF.  The second derivative and higher picks up on only shape distortions instead.  In this way, decomposing the CCF variations into different terms separates out changes due to dynamic shifts versus changes due to differences in shape.  Recreating the time series using only derivatives of two or higher will give CCFs with only shape-driven variations.  The effects of these shape changes can then be removed from the time series.  The coefficients of the shape-driven derivative terms (i.e., $k \geq 2$) can also be used as activity indicators, as they reflect the magnitude of CCF variations due to changes in shape.

This method is conceptually similar to the \scalpels\ method described in Section \ref{sec:scalpels}.  In this framework, the quotients ($c_i(v)$) of each observation's CCF over a reference CCF is modeled whereas in \scalpels\ the autocorrelation function of the CCF or spectrum is used.  For \scalpels, the autocorrelation function is intrinsically insensitive to transitional shifts.  For \ccfprime, the higher-order ($k \geq 2$) derivatives are insensitive to transitional shifts.  These higher-order derivatives and their coefficients in the linear model capture the variation in the CCF and the magnitude of the variation, much as PCA does for \scalpels.  The coefficients of the linear model can also act as an activity indicator (much as the amplitudes from the PCA are used for \scalpels+\glom).   As the \ccfprime\ method remains exploratory, more work needs to be done to establish whether the different derivatives create an orthonormal basis as is the case with PCA.

The \ccfprime\ method requires only normalized CCFs and is straightforward to implement.  Higher resolution data will contain more information on the line profile distortions being modeled.  Higher SNR observations will give more accurate derivatives.  The observations should sample a broad range of activity states.  This ensures that changes in the CCF due to stellar signals are not reflected in the combined, reference CCF.  With many different manifestations of stellar signals in the range of CCFs, the specific features of any given activity state will be blurred out.

\subsection{\fiesta+\glom}\label{sec:fiesta}
The \fiesta\ method, submitted by the PennState team, decomposes the CCF of a spectrum into Fourier basis functions \citep{zhao2020,zhao2022}.  The shifts of each of these basis functions are then calculated for a range of Fourier frequencies.  A pure CCF shift will manifest as a constant shift in all Fourier frequencies and can easily be subtracted out.  Shape deformations, on the other hand, will be frequency dependent.  This decomposition therefore parameterizes the effects of stellar signals as a series of shifts at each frequency for each CCF. These frequency-dependent shifts can be used together as a multi-dimensional activity indicator.

The \fiesta\ method reads in CCFs for each observation.  These CCFs must be properly normalized as a vertical offset could also produce a frequency-dependent shift that would be mistaken for a shape deformation.  Observations with greater SNR allow for more frequencies to be incorporated.

The activity indicators produced by \fiesta\ were post-processed using principal component analysis \citep{zhao2022} and modeled jointly with dynamical RV shifts using GLOM (as described in Section \ref{sec:glom}).

\subsection{CCF Linear Regression}\label{sec:ccfLR}
The CCF Linear Regression method, submitted by the \mleprv\ team, makes use of machine learning to model variations in the CCF that are expected to be due to stellar signals \citep{debeurs2020}.  Specifically, the machine learning model predicts the difference between a Gaussian fit to the CCF and the true velocity shift.  This prediction can then be subtracted from the input RVs to give corrected RVs.

This method requires CCFs for each exposure and best fit RVs.  The CCFs are first shifted by the best-fit RVs so there are no translational difference between the different CCFs.  This allows the model to instead focus on shape variations.  The model is fed differential CCFs, i.e., the residuals from subtracting a reference CCF (made by taking the median of all CCFs) from each CCF.  These differential CCFs are normalized by the median and standard deviation of each point in the CCF across all observations such that the variations are roughly equal in magnitude.

In order to reduce the complexity of the model, only about four to six locations across the residual CCFs are modeled using a linear regression model.  The more observations there are, the more locations can be used without the risk of over fitting.  The base model for a single CCF and associated RV is given by:
\begin{equation}
RV = w_1\cdot CCF_1 + w_2\cdot CCF_2 + \cdots + w_v\cdot CCF_v
\end{equation}
where $CCF_v$ is the value of the differential CCF at velocity $v$ and $w_v$ is the associated weight parameter that is fit for.

Two slightly more complicated models were also tested.  For all targets, \ha\ information was added to the model to give:
\begin{equation}
RV = w_1\cdot CCF_1 + w_2\cdot CCF_2 + \cdots + w_v\cdot CCF_v + b\cdot H\alpha
\end{equation}
where \ha\ is the derived \ha\ emission for the given exposure and b is the associated weight that is fit for like the $w_v$ weights are.  For HD 26965, a fitted Keplerian was also added with a fitted weight parameter $d$ as follows:
\begin{equation}
RV = w_1\cdot CCF_1 + w_2\cdot CCF_2 + \cdots + w_v\cdot CCF_v + b\cdot H\alpha + d\cdot Keplerian
\end{equation}

Each of the CCF Linear Regression model versions included several measures to prevent over-fitting.  Specifically, the method results can be very sensitive to the choice in location across the differential CCFs.  To address this concern, the implementation (1) used significantly less free parameters than observations (i.e., four to free parameters for 25 to 58 observations).  (2) The magnitude of the weights for each CCF location was limited given that large weights are a common sign of over-fitting.  (3) CCF locations were checked to ensure they are capturing the general behavior in shape changes around that location rather than over-fitting.  This was done by shifting all CCF locations in x and seeing whether the results were comparable to shifting one CCF location at a time.  In the future, implementing a cross-validation approach would further address over-fitting concerns.

This CCF Linear Regression method does not use timing information.  Though it benefits from more observations, the cadence of these observations does not matter.  More observations allow for more locations in the differential CCFs to be included in the model, allowing it to potentially pick up on more shape variations.  The method can be sensitive to choice of locations across the differential CCFs, which require some fine-tuning.

\section{In-Depth Descriptions of Line-by-Line Methods}
\subsection{CCF Mask-\vald}\label{sec:wiseCcf}
The CCF Mask-\vald\ method, submitted by the PennState team, aims to generate cleaner CCFs by mitigating the effects of variable lines, blended lines, telluric contamination, and lines strongly affected by stellar variability and activity.  First, an automatic line-fitting code finds all spectral lines and fits them to a Gaussian with a linear offset.  Fitted line depths are used as mask weights for each line.  Any spectral line with a line center falling withing 30 \kms\ of features in the provided \selenite\ telluric model were removed.

A line list from the Vienna Atomic Line Database (\vald) is used to vet lines too near each other in order to avoid line blends.  For each target, an optimal definition of ``too near" was empirically determined, where any lines with centers closer than a given line blend cutoff were removed.  Cutoffs ranging between 0 to 27 \kms\ in intervals of 3 \kms\ were tested.  Masks used a Gaussian window function.  Different mask widths were tried where the sigma of the Gaussian window function ranged from one to eight pixels.  The optimal mask window width and line blend cutoff was decided by the combination that gave the lowest resultant RV RMS.

Generating these masks requires the spectra along with a telluric model.  The approximate RV shift of each spectra as well as the expected line velocity width makes line-fitting easier.  The target star's stellar temperature and $\log g$ are needed for the \vald\ line list.

\subsection{CCF Mask-BIS and CCF Mask-RV}\label{sec:warwickCcf}
The CCF Mask-BIS and CCF Mask-RV methods, submitted by the Warwick team, constructs weighted, binary masks to remove the contributions from blended lines or lines particularly sensitive to stellar signals \citep{lafarga2020}.  Spectral lines are found by identifying relative minima in a high SNR stellar template built by coadding observations. Each line is then parametrized by fitting a Gaussian function.  This gives an initial line list with rest wavelengths for all lines.  Only lines with widths, depths, and asymmetry that fall between a specified range (as specified in \citet{lafarga2020}) are kept.  This ensures that the included lines are clear, sharp lines with no obvious blends. The provided \selenite\ telluric model is used to vet for any lines too near a telluric feature.

RVs are then computed for each individual line in each of the observations.  Each line is fit to a Gaussian.  The mean of this Gaussian is taken to be the line center, which is then compared to the initial line list to calculate the RV shift of the line.  Lines are determined to be either sensitive or insensitive to photospheric velocities based on how correlated they are with a given activity indicator.  The Pearson correlation coefficient is used to gauge the degree of correlation.  Lines were established as inactive if they had a coefficient less than 0.2-0.4 and spread in RVs less than 10-15 \ms\ (with the specific cutoff depending on the target).  Active lines had correlation coefficients greater than 0.3-0.5 with RVs or a correlation coefficient less than or equal to -0.3 in the case of the BIS-guided mask.

Very correlated lines are likely to be strongly affected by stellar signals.  If a line's RVs exhibit a lot of scatter, it becomes difficult to tell whether a line is truly uncorrelated with an activity indicator, or if the correlation is merely lost among the scatter.  Therefore, lines that exhibit a large RV scatter are also discarded.  The remaining lines that exhibit little to no correlation with activity indicators are averaged to compute a final RV for each exposure.

The results presented in this report used either the CCF BIS (CCF Mask-BIS) or the CCF RV (CCF Mask-RV) as an indicator to establish what lines are strongly correlated with stellar signals.  Note, the CCF RV and individual line RV are not fully independent, which could bias the correlations measured.  Other than choice of indicator, there is no specific tuning required for this method.

For this method, the data must be high enough resolution to resolve line blends.  The data should also be stable enough that the dominate variations in lines are due to stellar signals and not instrumental or other non-astrophysical effects.  More observations, especially over a greater range of activity states, will result in a better measure of correlation.

\subsection{\lblspec, \lblrvel, and \lblboth}\label{sec:lblrvs}
The Geneva team used a combination of spectral cleaning techniques and line-by-line RVs.  The provided spectra were first continuum normalized using \rassine, an open source python package that makes use of convex hulls to determine continuum points \citep{xu2019, cretignier2020-02}.  \yarara\ was then used to clean the spectra of tellurics and first-order morphological variations away from a median spectra \citep{cretignier2021}.  Using this post-processed spectra, a master spectrum and tailored stellar mask (to avoid line blends) was developed for each star.

Line-by-line RVs were extracted, where RVs for each spectral line are derived relative to the star-specific master spectrum \citep{dumusque2018}.  With \lblspec, a weighted PCA is run on the spectral level and the first three components are used to reconstruct a denoised, master spectrum.  The degree to which lines are affected by stellar signals or observational systematics varies from line to line, as reflected in the spread of each line-specific RV across all observations.

For \lblrvel, PCA is used to identify variations across all lines in all observations, where each observation has been corrected by its average RV.  The first three principal components are used to decorrelate the average RV signal for each observation using a multi-linear regression.

This method is run using merged spectra, where all echelle orders of a spectrum have been merged to form one, long spectrum.  The basic method described here requires little tweaking to run, but implementing \yarara\ can get increasingly more complex if it is used to do a more tailored job of removing instrumental systematics.  Because each line now stands alone, this analysis does require higher SNR spectra in comparison with a classic CCF.  In order to use \yarara\ to disentangle telluric features, the input set of observations must have a good coverage of different barycentric shifts in order to separate the stellar lines from the telluric lines.  For best performance from the PCA, it is ideal if the observations also cover a wide range of stellar activity states.

This method outputs RVs for every line as well as the principal variation in the centered RVs from the RV-level PCA.  The PCA here is run directly on the line RVs or the spectra itself rather than chromospheric proxies, such as more classic activity indicators.  The PCA step might be swayed by outliers or the presence of large variation, e.g., hardware changes, abnormal observing conditions, etc.  By using the whole spectrum and treating each line independently, LBL RVs reveal how individual lines are affected by variations from either stellar signals or instrument systematics.  This gives a better picture of how these affects are manifesting in the spectra.

There are three flavors of LBL results presented here.  The \lblspec\ results uses PCA at the spectral level to create the master tempalte while the \lblrvel\ method implements PCA on the recovered RVs for each line.  Both methods can also be combined by first applying the PCA decomposition to the spectra, extracting LBL RVs using that master template, and then decomposing the resultant LBL RVs with another PCA.  Those results are included as \lblboth.

\subsection{\pwgp}\label{sec:pwgp}
The Pairwise Gaussian Process RV Extraction method, submitted by the \oxgen\ team, uses GPs to model and then align all pairs of spectra with each other \citep{rajpaul2020}.  These pairwise RVs can then be combined to establish differential RVs without having to construct a master template.  The pairwise matching is done on a highly localized basis---i.e., each spectra is broken up into many different ``chunks" with each chunk containing one to a few spectral features.

These smaller chunks can be treated as independent measures of the spectral shift, where some chunks will contain more RV information or be more affected by stellar variability than others.  More sophisticated implementations are possible, for example modifying the GP modeling of spectral chunks to model stellar variability in addition to Doppler shifts.  For the results presented here, spectral chunks that appeared ``contaminated" by stellar variability were simply not used when computing final RVs.

The \pwgp\ method reads in spectra.  A Mat\'{e}rn $\frac{5}{2}$ kernel is used to model and align each spectral chunk, with different hyper-parameters returned for each chunk.  This can get quite computationally expensive, but is helped by the pairwise framework.  Though the method requires little tuning to run, some thought must go into deciding which chunks are considered ``contaminated" and what to do with them.  

There are many possible metrics to use in determining which chunks appear to be contaminated.  The chunk itself may exhibit unusually large variation from one exposure to another, suggesting there are stellar signals or tellurics present in the chunk that is causing it to return such a large range of RV measurements.  Similarly, the RV error of a chunk may be higher than typical.  The RVs of a chunk may also show statistically significant correlation with an activity indicator, suggesting the RV from that chunk is mostly due to stellar signals rather than true dynamical shifts.

Tuning the cut offs for which chunks to include requires balancing between the RMS of the final RVs and the error bars on these measurements.  Removing too many chunks will exclude too much data from the process, thereby increasing the error bars for each RV measurement.  Not removing enough chunks means noise will continue to be incorporated into the final RV measurements, thereby resulting in greater RV scatter.

After cutting contaminated chunks, the RV measurements of the remaining chunks are combined to recover final RVs.  The RV from each chunk is inversely weighted by the scatter in returned RVs for that chunk as determined via a Markov chain Monte Carlo (MCMC) analysis.  By using a MCMC, the resultant weight incorporates both the photon noise and uncertainty from the GP fit.

Using GP modeling to align spectra should perform better (as compared to non-GP models) with lower-resolution and lower-SNR spectra.  However, having higher SNR/resolution spectra is needed when identifying contamination.

This method benefits from using a principled, GP modeling framework for spectral interpolation and alignment.  This precludes the need to generate a master template and indeed does not require any information about where lines are, what they may look line (i.e., depth, width, etc.), or how they might change with stellar signals.  On the other hand, the model also can not incorporate any prior knowledge of stellar or telluric contamination and does not distinguish between different forms of contamination whether stellar, terrestrial, or instrumental.

\section{In-Depth Descriptions of Methods That Model the Spectra}
\subsection{\dcpca\ and \dcpca+\glom}\label{sec:dcpca}
The Doppler-Constrained Principal Components Analysis method, submitted by the PennState team, identifies the largest variations in RV shifted spectral data using PCA \citep{jones2017}.  The resultant principal components highlight where the spectra is changing the most while the corresponding amplitudes of each principal component captures the magnitude of this change for each observation.  By feeding the PCA the full spectral format, the PCA is able to pick up on changes at the pixel level.  The principal component amplitudes can be used as an activity indicator.
  
The \dcpca\ method requires spectra and initial guess RVs for each observation.  The spectra are first shifted by the best-fit RV for each observation and then interpolated onto a common wavelength grid using a GP with a Mat\'{e}rn $\frac{5}{2}$ kernel.  Some tuning of what parts of the spectra to include in the PCA will help ensure the PCA is not picking up on variations from the instrument or tellurics.  While the method can be run on the full spectrum, the results reported here used the areas of spectra near lines specified by a CCF mask.  This helps to avoid telluric contamination and blended lines.

The number of principal components to incorporate into the analysis can be chosen in a number of ways.  As always, only principal components with significant features (i.e., are not purely noise) should be used.  With enough exposures, a classic cross-validation test can be used to gauge the performance of incorporating different numbers of components.  More observations will likely result in more significant components.  A component can also be tied to photospheric velocities if the amplitudes of the component are correlated with activity indicators.  Data with a high SNR and high resolution makes variations in the spectra clearer.  A broad wavelength coverage would also help, as it would encompass more changes. 

For the results presented in this report, the amplitudes of the first two principal components were used as indicators.  The publically available \espresso\ masks (also used to generate the provided CCFS) were used to determine which segments of the spectra were fed into the PCA.  The RVs were decorrelated against the resultant principal component amplitudes via both a simple linear regression and using the \glom\ framework with the sum of two Mat\'{e}rn $\frac{5}{2}$ kernels.

\subsection{\horatio\ and \horatio\ Self}\label{sec:horatio}
\horatio, submitted by the CCA team, takes the residuals of each observed spectrum against a Doppler-shifted template spectrum and regresses these residuals against housekeeping data, such as provided RVs, activity indicators, or instrumental measurements.  \horatio\ is so named as it operates under a generative framework---it constructs a model using a finite number of housekeeping data sets, or labels, to predict what the residuals will look like.  In doing so, \horatio\ establishes what properties of the residuals can be tied to the different effects being traced by the housekeeping data, be it stellar signals, instrument systematics, or whatever else it is given.  These effects that are not due to an orbiting planet can then be removed.

For $N$ observations, let $F$ represent all residuals from a model for each pixel of each spectrum while $Q$ represents all housekeeping data being used including the RVs.  We use $\Delta f_n$ to denote the residuals of a given observation $n$ and $\hat{q}_n$ to represent the predicted RV correction for that observation.  For a statistically rigerous model, for each observation $n$ or validation set, the $\Delta f_n$ residuals should be left out of $F$.  RV corrections can then be calculated as follows:
\begin{equation}
    \hat{q}_n = \Delta f_n \frac{dF}{dQ} \cdot \left[ \frac{dF}{dQ} \cdot \frac{dF}{dQ}\right]^{-1}
\end{equation}
where $\frac{dF}{dQ}$ represents the spectral residuals being regressed against the housekeeping data.  This is a first-order regression model.  The housekeeping data can vary depending on what is needed to give a complete, orthogonal representation of the variations being modeled.

Implementing this method requires spectra of each observation and housekeeping data associated with each spectra.  The template spectrum can be generated in any number of ways.  Higher resolution spectra will preserve more evidence of stellar variability in the residuals.  The regression itself is computationally simple to implement.

For the results presented in this report, a model spectrum was generated using \wobble, a data-driven method for extracting RVs and inferring the underlying spectral components \citep{bedell2019}.  The CBC RVs and \ha\ equivalent width are the housekeeping data used.  Expected RV offsets are calculated using a cross-validation framework where an eighth of the data at a time is left out of the model construction.  For reference, the results where all data is used is given as \horatio\ Self results.  For both the cross-validation and self frameworks, all observations are used to construct the model spectrum with \wobble.

By incorporating all the spectral residuals, \horatio\ is able to incorporate information from every pixel of the spectral data.  The housekeeping data is then used to try and predict the behavior of different pixels and the magnitude of change to the RVs expected from these variations.  Incorporating more data that traces different effects makes the method more sensitive to different causes of spectral variations.  On the flip side, the method is also incapable of tracing any variation not associated with the provided housekeeping data.  The regression will be poorly constrained if the housekeeping data sets used are not all independent and do not all trace a real change on the residuals being modeled.

\subsection{\hamlet}\label{sec:hamlet}
\hamlet, submitted by the CCA team, is similar to \horatio\ and also regresses spectral residuals to a shifted template against housekeeping data.  \hamlet, the discriminative counter part to \horatio, is discriminative in that it uses the residuals to predict the housekeeping data.  The result is a prediction of the magnitude of RV shift due to observation-specific spectral variations as captured in the residuals to a spectral model.

As with \horatio, let $F$ represent the array of all spectral residuals, $\Delta f_n$ the residuals for a given observation $n$, and $Q$ be the array of RVs acting a labels.  The predicted RV correction for each observation, $\hat{q}_n$, can then be calculated
\begin{equation}
    \hat{q}_n = \Delta f_n \cdot \left( F^TF+\alpha I \right)^{-1}F^TQ
\end{equation}
where $\alpha$ represents an opportunity to introduce expected information content, for example uncertainties on the spectral residuals or spectral resolution.

The inputs, implementation, and output for the \hamlet\ method is the same as for the \horatio\ method described above.  After acquiring the residuals to a template spectra and associated RVs for each spectra, the method takes seconds to run.  The only housekeeping data used for \hamlet\ are the CBC RVs for each exposure.

The discriminative framework is more agnostic about precisely what housekeeping data is included.  The regression itself works to construct an orthogonal transformation that can be mapped onto the derived RVs.  This framework is more appropriate in the regime where the spectra is varying in more ways than can be captured by the provided housekeeping data.  Since it is not clear whether known activity indicators trace all possible spectral variations due to stellar signals, the discriminative framework may be more appropriate than the generative framework for disentangling photospheric velocities from true center-of-mass shifts.

In truth, there is a latent model that produces both the housekeeping data and the spectral variations, namely the activity and intrinsic variability of the target stars.  Both the generative and discriminative frameworks move between the products of this latent model, just in different directions.  Both the \horatio\ and \hamlet\ methods are ongoing work; the results presented here are an initial implementation of the two methods.

\section{Submitted RVs of All Methods} \label{sec:rawResults}
The following section show the submitted RVs, both clean and activity RVs where available, as well as their periodograms.  Methods are presented (top-to-bottom, left-to-right) in the order in which they are presented in the Methods Section and Appendix.  Given the large format of the figures, their content is described here in the text.

The top-left plot shows the originally provided \expres\ RVs (first column) along with the periodogram (second column) in black and the periodogram of the time sampling, or the window function, in green.  The rest of the rows show the submitted clean RVs in blue.  Each figure is labeled by the team and method name.

The periodogram subplots for each method shows a periodogram of the clean RV in blue.  If provided, the periodogram of submitted activity RVs are also shown in orange.  A significance level of p-value = 0.01 is shown as a horizontal, black line across the periodograms.  A p-value of 0.1 is shown as a dashed black line. Axes with the words ``No Submission" are shown for methods that did not submit results for that target.

\begin{figure*}[h]
\centering
\includegraphics[width=.9\textwidth]{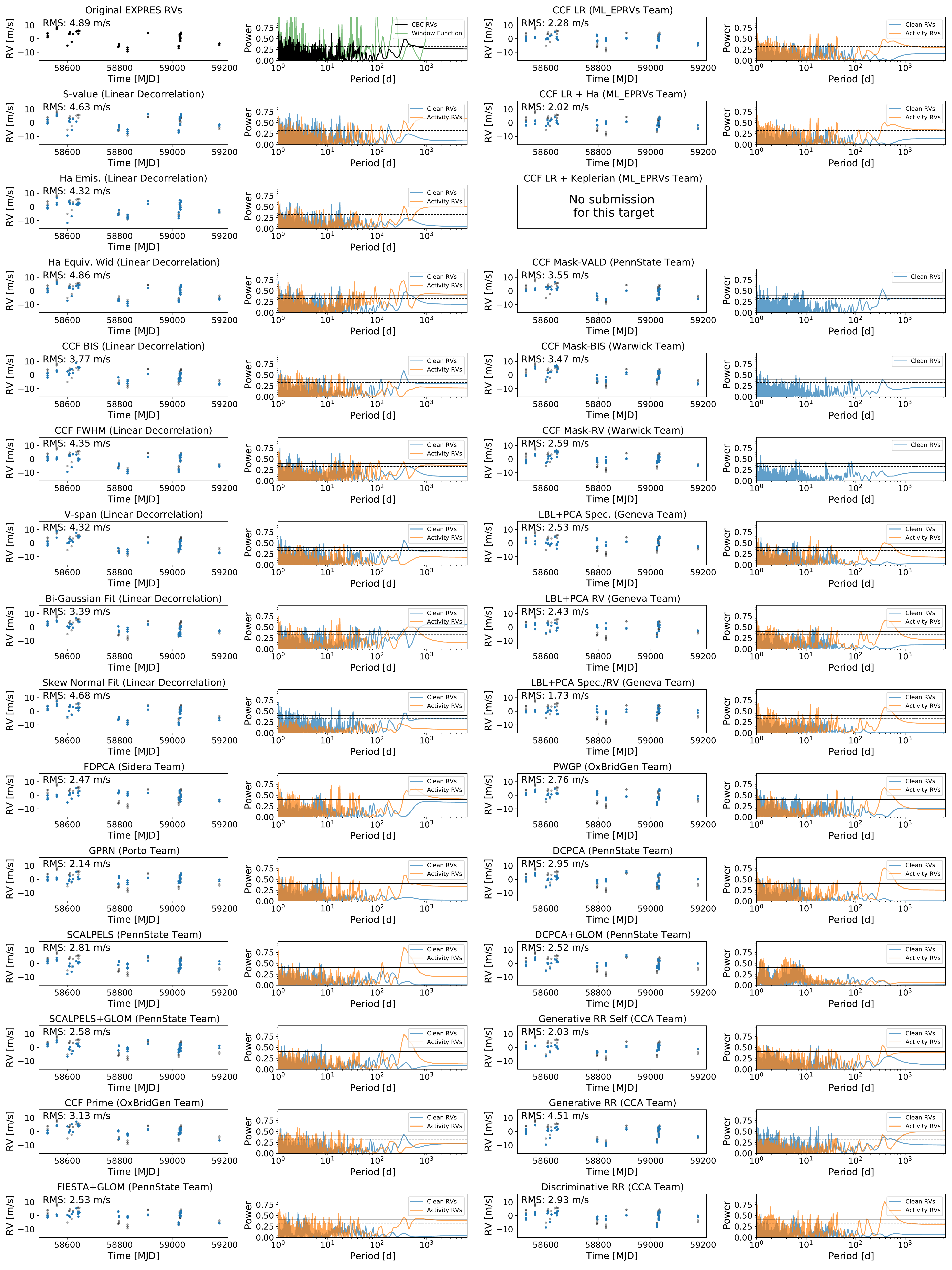}
\caption{Submitted results for HD~101501.  For each periodogram, p-values of 0.01 and 0.01 are shown as horizontal solid and dashed black lines respectively.}
\label{fig:101501}
\end{figure*}

\begin{figure*}[h]
\centering
\includegraphics[width=.9\textwidth]{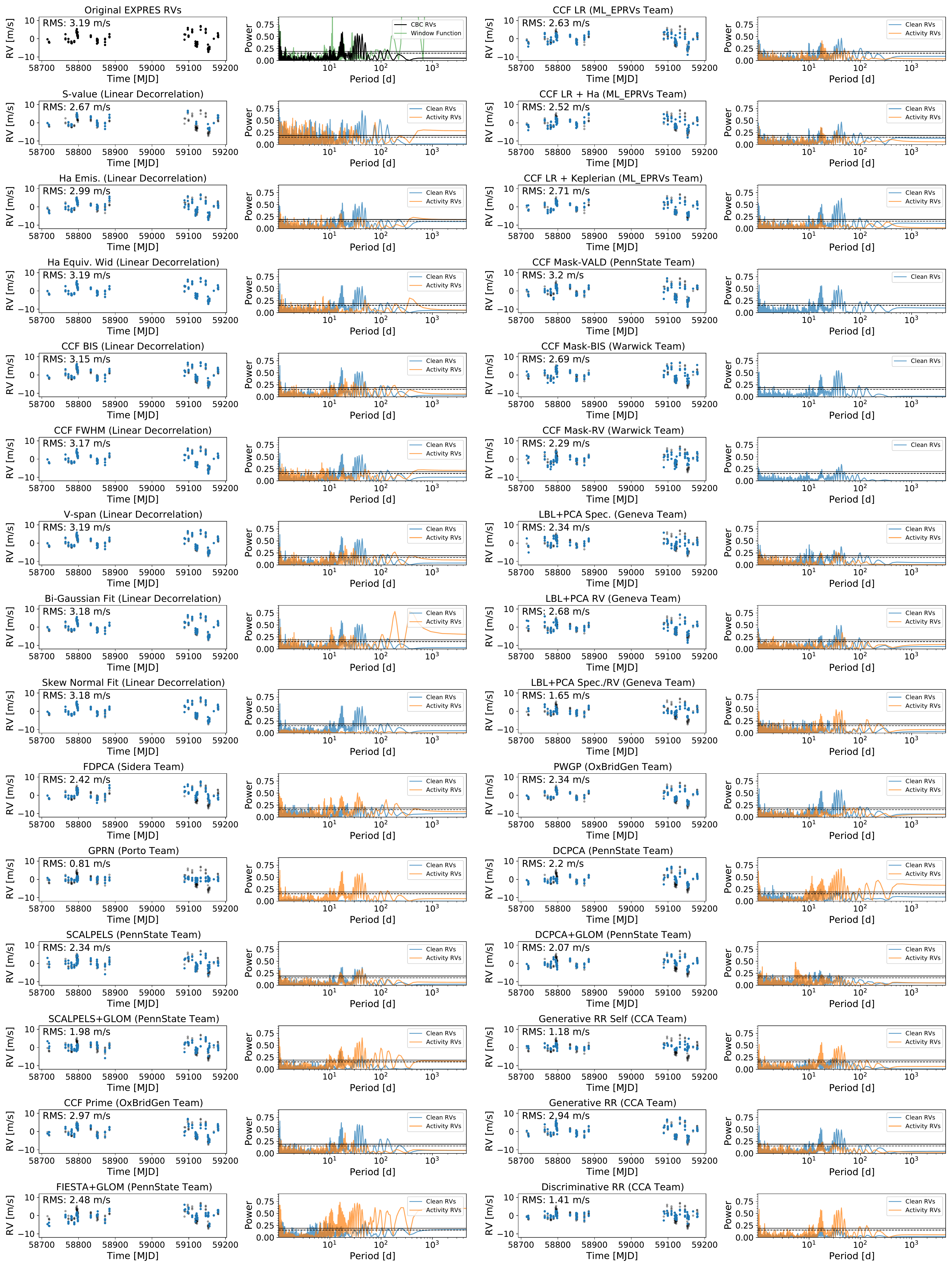}
\caption{Submitted results for HD~26965.  Otherwise the same as Figure \ref{fig:101501}}
\label{fig:26965}
\end{figure*}

\begin{figure*}[h]
\centering
\includegraphics[width=.9\textwidth]{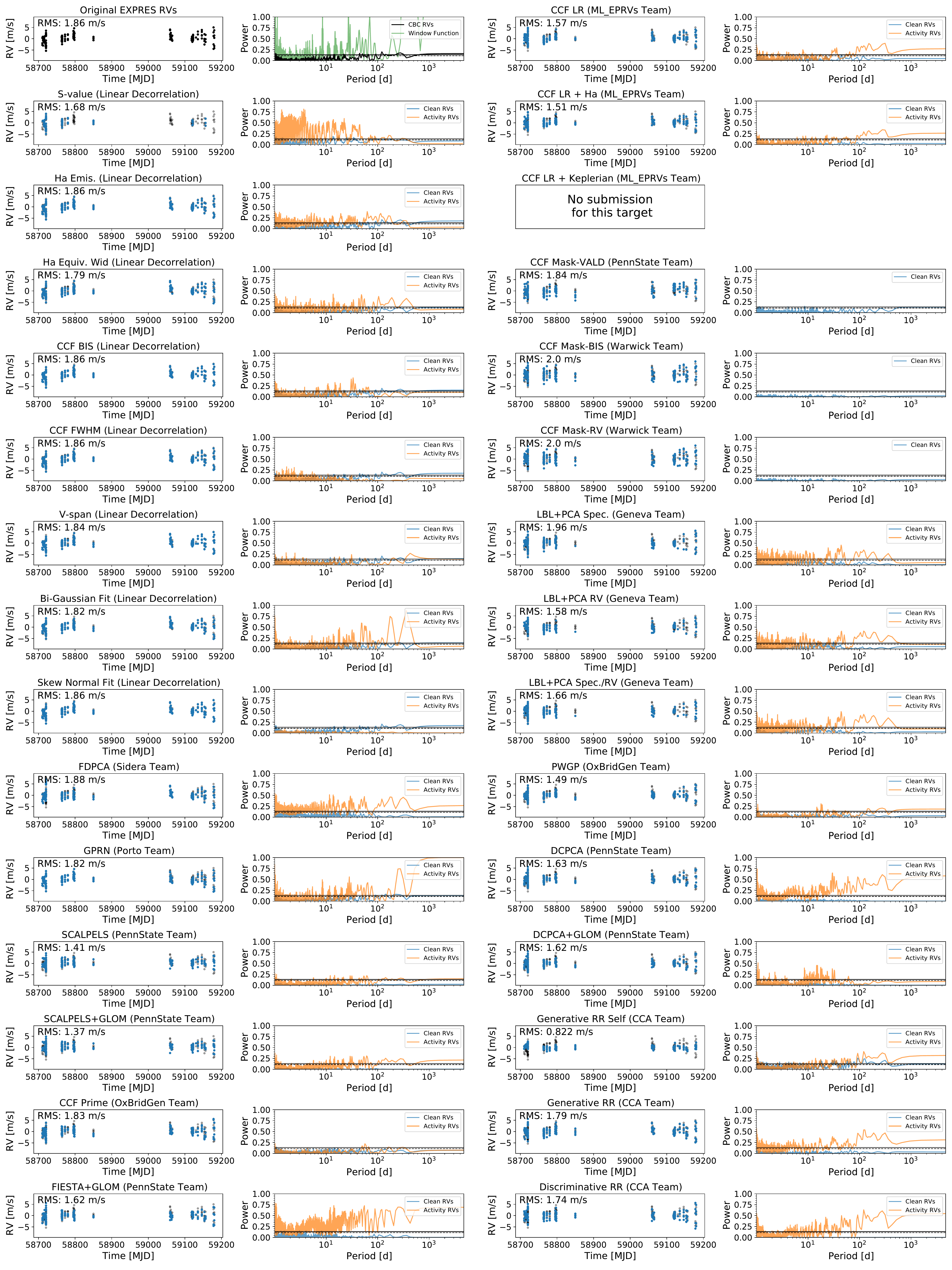}
\caption{Submitted results for HD~10700.  Otherwise the same as Figure \ref{fig:101501}}
\label{fig:10700}
\end{figure*}

\begin{figure*}[h]
\centering
\includegraphics[width=.9\textwidth]{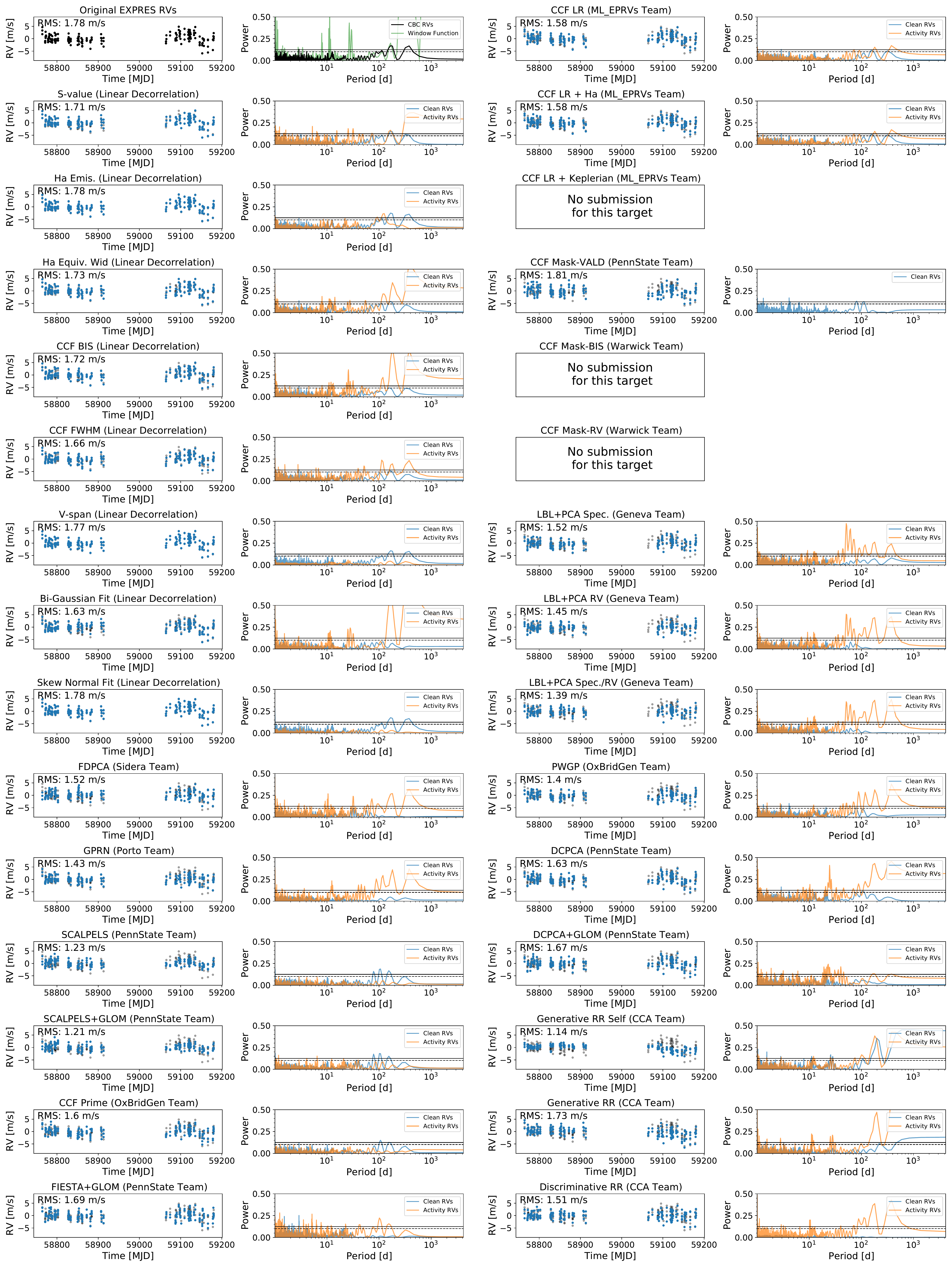}
\caption{Submitted results for HD~34411.  Otherwise the same as Figure \ref{fig:101501}}
\label{fig:34411}
\end{figure*}

\end{document}